\documentclass[seceq]{ptptex}

\def\>{\rangle}
\def\<{\langle}
\def\ket#1{|#1\>}
\def\bra#1{\<#1|}
\def\braket#1#2{\< #1 | #2 \>}
\def\bracket#1#2#3{\< #1 | #2 | #3 \>}
\def\ave#1{\< #1\>}
\def\aave#1{\<\!\< #1 \>\!\>}
\def\ve#1{{\bm{#1}}}
\def\ma#1{{\rm \mathbf{#1}}}
\def\tr{\,{\rm tr}\,}
\def\one{\mathbbm{1}}
\newcommand{\op}[1]{\hat{#1}}

\usepackage{bbm}
\usepackage{bm}
\usepackage{graphicx}
\usepackage{wrapft}

\title{Theory of quantum Loschmidt echoes}

\author{Toma\v z \textsc{Prosen}$^1$, Thomas H. \textsc{Seligman}$^2$ and
Marko \textsc{\v Znidari\v c}$^1$}

\inst{${}^1$ Physics Department, FMF, University of Ljubljana, Ljubljana, 
Slovenia\\
${}^2$ Centro de Ciencias F{\'\i}sicas, University of Mexico (UNAM), 
Cuernavaca, Mexico}

\abst{
In this paper we review our recent work on the theoretical approach to
quantum Loschmidt echoes, i.e. various properties of the so called echo
dynamics -- the composition of forward and backward time evolutions generated by 
two slightly different Hamiltonians, such as the state autocorrelation function
(fidelity) and the purity of a reduced density matrix traced over a subsystem 
(purity fidelity).

Our main theoretical result is a linear response formalism, expressing the
fidelity and purity fidelity in terms of integrated time autocorrelation function
of the generator of the perturbation.
Surprisingly, this relation predicts that the decay of fidelity 
is the slower the faster the decay of
correlations. In particular for a static (time-independent) perturbation, and
for non-ergodic and non-mixing dynamics where
asymptotic decay of correlations is absent, a qualitatively different and faster decay
of fidelity is predicted on a time scale $\propto 1/\delta$ as opposed to
mixing dynamics where the fidelity is found to decay exponentially on a time-scale 
$\propto 1/\delta^2$, where $\delta$ is a strength of perturbation.
A detailed discussion of a semi-classical regime of small effective values of
Planck constant $\hbar$ is given where classical correlation functions can be used to 
predict quantum fidelity decay. Note that the correct and intuitively 
expected classical stability behavior is recovered in the classical limit 
$\hbar\to 0$, as the two limits $\delta\to 0$ and $\hbar\to 0$ do not commute.
The theoretical results are demonstrated numerically for two models, the quantized kicked top and 
the multi-level Jaynes Cummings model. Our method can for example be applied to the stability 
analysis of quantum computation and quantum information processing. 

}

\begin{document}

\maketitle

\section{Introduction}

In the early days of statistical mechanics, J. Loschmidt in a discussion with L. Boltzman
suggested to study irreversibility by changing the velocities of all the molecules in a box
which are, after an equal amount of time, supposed to return to their initial positions.
Since the discovery of deterministic chaos, however, we know that such a 
{\it gedanken} experiment 
won't work after a short while since typical systems of many particles
possess exponentical sensitivity on the variation of initial condition.
Therefore arbitrarily small but non-vanishing perturbation of the trajectory (i.e. 
initial condition), or perturbation of the equations of motion (e.g. introducing a small external
force field like gravity), drive the returning orbit away and hence the orbit will
never return to the initial phase space point. Both mechanisms, namely perturbing the 
initial condition or perturbing the Hamiltonian, produce a similar effect in classical mechanics.

This, however, is not the case in quantum mechanics. Due to manifest linearity and unitarity of quantum 
equation of motion (Schr\" odinger equation), quantum dynamics is always stable against 
small variations of the initial state described by the wave-function \cite{casati86}. 
Yet small variation in the 
Hamiltonian can produce interesting and highly non-trivial effects on quantum time evolution.

It has been suggested by A. Peres \cite{Peres1,Peres2} that susceptibility of quantum evolution to 
small system variations, as measured by the Fidelity $F(t)=|\braket{\psi(t)}{\psi_\delta(t)}|^2$ of 
the states of the perturbed $\ket{\psi_\delta(t)}$ and unperturbed $\ket{\psi(t)}$ time evolution, 
can provide a useful signature of classical chaos in quantum motion.
Peres suggested that  classically chaotic systems are characterized by fast exponential decay of 
fidelity $F(t)$, while the decay in regular systems should be qualitatively slower. 
This conclusion was asserted, based on
a numerical experiment where an initial coherent state has been placed, respectively, inside the
chaotic region of classical phase space, or in the {\em middle} of KAM island of classically 
regular motion. This view, elaborated semiclassically by Jalabert and Pastawski \cite{Jalabert} 
in the regime of initial coherent states and very short time scale, i.e. shorter than the so-called 
Ehrenfest time $t_{\rm E} \propto -\log\hbar$ \cite{Berman}, is consistent with a semiclassical picture of 
decoherence of Zurek \cite{Zurek}, which predicts, in the same {\em semiclassical regime}, that von Neumann 
entropy of a reduced density matrix of a central system traced over the environment grows with the 
rate proportional to the classical Lyapunov exponent (or better to say, local classical phase space 
stretching rate). According to this picture, fidelity decays exponentially $F(t) \propto \exp(-\lambda t)$
for classically chaotic systems, with the {\em perturbation independent} rate $\lambda$ 
which matches the local classical phase-space stretching rate.

We stress that this picture is justified under two rather severe assumptions: (i) coherent initial states which
allow quantum-classical correspondence in phase space and (ii) short-times $t < t_{\rm E}$ which
guarantee pointwise quantum-classical correspondence of time evolution. If either of the two conditions
is broken, the above picture can be shown to be incorrect \cite{ktop}.
In quantum information processing, in particular, one is certainly not interested in processing coherent 
initial states but rather {\em random} initial states which contain a maximal amount of quantum information.
Furthermore, one is interested in the behaviour of fidelity $F(t)$ for asymptotically long times,
certainly longer than the Ehrenfest barrier $t_{\rm E}$.

In a series of papers \cite{Prosen01,QC,ktop,KIJPA,JCPRL,ktop2} 
we have elaborated on a linear response approach to fidelity decay which allows to deduce 
the behaviour of $F(t)$ in the entire range
of times for sufficiently small perturbation strength. In several cases linear response (perturbative)
results can be extended to include all-orders in perturbation parameter and yield asymptotic 
long-time tails of $F(t)$. The central result of this work is a linear-response fluctuation-dissipation
formula which expresses fidelity $F(t)$ in terms of integrated time-correlation function of the 
perturbation. As a general consequence of this formula, it follows that stronger correlation decay
(typically associated with stronger classical chaos of the underlying classical counterpart)
means higher fidelity, or slower decay of fidelity. This has been quite an unexpected result as
it seems just oposite to the short-time semiclassical picture \cite{Jalabert}. However it can be
shown, as a consequence of a delicate competition of time-scales (see e.g. Ref.\cite{ktop} or the present paper), 
that the two pictures are not contradictory but they're in fact complementary, and that 
there is a crossover of fidelity decay $F(t)$ at $t\sim t_{\rm E}$ from classical to quantum behaviour.

Fidelity decay is sometimes associated with decoherence, which is a dynamical property of open
quantum system coupled to another quantum system interpreted as environment. In order to elucidate 
this from a dynamical point of view, we have introduced \cite{KIJPA,JCPRL,ktop2} a novel concept of 
{\em purity fidelity}, namely the purity of a reduced density matrix of a central system traced 
over the environment (or another part of a system) after undergoing the (Loschmidt) echo dynamics. We show 
an intimate relationship between purity fidelity and fidelity decay, and develop a linear 
response theory for purity fidelity in terms of special time-correlation functions of the generator
of perturbation.

This paper is a short comprehensive review of theoretical results on fidelity and purity fidelity
which appeared in a series of recent papers \cite{Prosen01,QC,ktop,KIJPA,JCPRL,ktop2}. 
Beyond that we discuss important aspects which have not yet been considered before, such as
the case of coherent initial states, regular classical dynamics and `ergodic' perturbation with
vanishing time-average.

It should be noted that a considerable recent interest in quantum Loschmidt 
echoes \cite{Jalabert,Prosen01,QC,ktop,KIJPA,JCPRL,ktop2,Beenakker,Pastawski,bruno,Tomsovic,benenti,veble,therest} has 
been largely stimulated by spin echo experiments performed by the group of 
H. Pastawski \cite{Usaj}, and that fidelity is used as a benchmark for the
quantum information processes.\cite{qcomp}

\section{General theory of fidelity: linear response and beyond}

Let us consider a unitary operator $U$ being either (i) a short-time propagator 
$U = \exp(-i H \Delta/\hbar)$ generated by some time-independent Hamiltonian $H$, 
or (ii) a Floquet map $U = \op{\cal T}\exp(-i\int_0^p d\tau H(\tau)/\hbar)$ of 
periodic time-dependent Hamiltonian $H(\tau+p)=H(\tau)$. 
Arbitrary time evolution is generated by a 
group $U^t$ where {\em integer} $t$ is a discrete time variable.
Note that in the continuous time case (i) we may let $\Delta \to 0$, so $\tau = t \Delta$ becomes
a continuous time variable. Thus we shall consider the case of discrete time 
and develop general formalism for that case, whereas in the end the formulae for 
continuous time shall simply be obtained as the limit $\Delta\to 0$.

A general (static, for time dependent case see Ref.\cite{QC}) 
small perturbation of unitary propagator $U$ can be `para\-met\-rized' in terms
of some bounded hermitian operator $V$ as
\begin{equation}
U_\delta = U\exp(-i V\delta/\hbar), 
\label{eq:U_d}
\end{equation}    
where $\delta$ is a small strength parameter.
We note that $V$ corresponds to the perturbation of the Hamiltonian $H_\delta = H + \delta V$ in
the continous time limit $\Delta\to 0$. We also want to keep dependences on the
 (effective) Planck constant $\hbar$ explicit so we can control the semiclassical behaviour.
For systems with a well defined semiclassical limit we shall be interested in the
perturbations $V$ which have well defined classical limit or Weyl symbol 
$v(\ve{q},\ve{p})$.

Starting from some initial reference state $\ket{\psi}$, we consider two quantum time
evolutions $\ket{\psi(t)} = U^t\ket{\psi}$, $\ket{\psi_\delta(t)} = U^t_\delta\ket{\psi}$,
and investigate a distance between the resulting Hilbert states as measured by
fidelity $F(t) = |\braket{\psi_\delta(t)}{\psi(t)}|^2$. It is crucial to note that fidelity can be written in terms of expectation value
$\ave{\bullet} = \bra{\psi}\bullet\ket{\psi}$ of the unitary {\em echo operator}
\begin{equation}
M_\delta(t) = U^{-t}_\delta U^t,
\label{eq:Mdef}
\end{equation}
namely
\begin{equation}
F(t) = |\ave{M_\delta(t)}|^2.
\label{eq:Fdef}
\end{equation}
We shall later discuss the dependence of fidelity on dynamical properties of 
the time-evolution $U$,
on the strength of perturbation $\delta$, and on the structure of initial state $\ket{\psi}$.

Concerning the last point it is often useful to study fidelity with respect to a {\em random} 
initial state $\ket{\psi}$, i.e. to average, denoted by $\aave{\bullet}$, over an ensemble of 
initial states with a unitarily invariant measure in $N$-dimensional Hilbert space.
This should generally correspond to processing of states with {\em maximal quantum information} 
(entropy) which is certainly the situation of most interest for quantum computation 
\cite{qcomp}. We note that bilinear expressions in $\ket{\psi}$ are averaged by means of a trace,
\begin{equation}
\ave{\bra{\psi}A\ket{\psi}}_\psi =: \aave{A} = \frac{1}{\cal N}\tr A,
\end{equation}
however higher order expressions, like fidelity (\ref{eq:Fdef}) which is of fourth order in random 
variable $\ket{\psi}$,
have to be averaged by means of pair contractions in the asymptotic Gaussian (${\cal N}\to\infty$) limit.
For example, one can compute the average fidelity $\aave{F(t)}$ in terms of the
average fidelity-amplitude $\aave{f(t)} = \aave{M_\delta(t)}$ with a small, semiclassically vanishing
correction
\begin{equation}
\aave{F(t)} = \left|\aave{f(t)}\right|^2 + \frac{1}{\cal N}.
\label{eq:1N}
\end{equation}
Thus we have shown that the variation of fidelity with respect to random 
initial states becomes unimportant for large Hilbert space dimensions ${\cal N}$, e.g. when
approaching either the semiclassical or the thermodynamic limit, so one is justified to average
the amplitude $f(t)$ in order to simplify theory.\cite{ktop}

On the other hand, for the purpose of quantum classical correspondence, it is most
suitable to consider {\em coherent} initial states $\ket{\psi}$, in some sense the states of
`{\em minimal quantum information}'. Below we shall specify our general formulae for the
two extremal cases, namely random and coherent initial states.

To derive our general theory we first rewrite the echo operator (\ref{eq:Mdef}) in terms of a
time-dependent perturbation operator in the {\em interaction picture}
\begin{equation}
V_t:=U^{-t}VU^t.
\end{equation}
Then we recursively insert the expression of unity $\one=U^{t'}U^{-t'}$ in the definition (\ref{eq:Mdef}) 
and observe $U^{-t'}U_\delta^{-1}U^{1+t'} = \exp(i V_{t'}\delta/\hbar)$
for $t'$ running from $t-1$ downto $0$, yielding
\begin{eqnarray}
M_\delta(t) &=& U_\delta^{-t} U^t = 
U_\delta^{-(t-1)}U^{t-1}U^{-(t-1)}(U_\delta^{-1} U) U^{t-1} = \nonumber \\
 &=& U_\delta^{-(t-1)}U^{t-1}\exp(i V_{t-1}\delta/\hbar) = \nonumber \\
 &=& U_\delta^{-(t-2)}U^{t-2}\exp(i V_{t-2}\delta/\hbar)\exp(i V_{t-1}\delta/\hbar) = 
\nonumber \\
 &\ldots&\nonumber \\
 &=& 
\exp(i V_0\delta/\hbar)\exp(i V_1\delta/\hbar)\cdots\exp(i V_{t-1}\delta/\hbar). 
\label{eq:Mprod}
\end{eqnarray}

The obvious next step is to expand the product (\ref{eq:Mprod}) 
into a power-series in $\delta$
\begin{equation}
M_\delta(t) = \one + \sum_{m=1}^\infty \frac{i^m \delta^m}{m!\hbar^m}
 \op{\cal T} \!\!\! \sum_{t_1,\ldots,t_m=0}^{t-1} V_{t_1} V_{t_2} \cdots V_{t_m},
\label{eq:Msum}
\end{equation}
where the operator $\op{\cal T}$ denotes a left-to-right time ordering. 
Such a perturbative expansion converges absolutely for any $\delta$ provided that the perturbation 
$V$ is a bounded operator. Therefore fidelity may be computed to arbitrary order in $\delta$
by truncation of the above expansion (\ref{eq:Msum}) and plugging it into (\ref{eq:Fdef}). 
So we can see that the fidelity $F(t)$ can be expressed entirely in terms of multiple time 
correlation functions of the generator $V$ of the perturbation. 

To second order in $\delta$ one obtains a very useful linear response formula
\begin{equation}
F(t) = 1 - \frac{\delta^2}{\hbar^2} \sum_{t',t''=0}^{t-1} C(t',t'') + {\cal O}(\delta^3),
\label{eq:F2nd}
\end{equation}  
where 
\begin{equation}
C(t',t'') := \ave{V_{t'} V_{t''}} - \ave{V_{t'}}\ave{V_{t''}}
\end{equation}
is a 2-point time correlation function of the quantum observable $V$.
This formula can be interpreted in terms of a dissipation-fluctuation relationship. On the LHS we have
fidelity which describes {\em dissipation of quantum information} and on the 
RHS we have an
integrated time-correlation function ({\em fluctuation}).
A simple-minded qualitative conclusion drawn from the formula 
(\ref{eq:F2nd}) says:
The stronger the decay of correlations the slower the decay of fidelity and vice versa.
The linear response formula can be rewritten in a slightly more compact notation as
\begin{equation}
F(t) = 1 - \frac{\delta^2}{\hbar^2} C_{\rm int}(t),
\label{eq:F2ndb}
\end{equation}
where
\begin{equation}
C_{\rm int}(t) = \ave{\Sigma(t)^2} - \ave{\Sigma(t)}^2,\quad
\Sigma(t) = \sum_{t'=0}^{t-1} V_{t'}.
\end{equation}
We shall use this compact notation later in section \ref{sec:pf}.

Note that the range of validity of the above formula 
(\ref{eq:F2nd},\ref{eq:F2ndb}) is {\em not} at all restricted to short times.
The only condition is that $\delta$ is 
sufficiently small such that $1-F(t)$ is small ($\ll 1$). Below we shall discuss 
different regimes and different time-scales based on this formula and its higher order 
extensions. In the semiclassical regime of approaching the classical limit $\hbar\to 0$ 
the quantum correlation function $C(t',t'')$ goes over to the corresponding correlation 
function of the classical perturbation $v(\ve{q}(t),\ve{p}(t))$.

Therefore the fidelity for classically chaotic systems will decay with the rate which is
{\em inversely proportional} to their rate of mixing. Furthermore for classically non-ergodic, 
i.e. regular or integrable motion, the correlation functions will generally not decay to zero 
and the fidelity will therefore decay much faster. 

For continuous time $t$ one expects that in the so-called Zeno regime
of short times, such that correlation function does not yet decay appreciably.
Fidelity will always decay quadratically $F(t) \approx 1 - \delta^2 C(0) t^2/\hbar^2$.
This is independent of the nature of the corresponding classical dynamics, whether it is 
regular or chaotic. However we are interested in longer times, beyond the range of quantum 
Zeno dynamics.

\subsection{Regime of ergodicity and fast mixing}

Here we assume that the system is (classically) ergodic and mixing such that the
correlation function $C(t',t'')$ decays sufficiently fast as $|t'-t''|$ grows; this typically 
corresponds to globally chaotic classical motion.
This also implies that after a certain short (Ehrenfest) time scale $t_{\rm E}$, 
which can for a chaotic system be written $t_{\rm E} \approx \log(1/\hbar)/\lambda$ in terms of
an effective classical Lyapunov exponent $\lambda$, the time correlation functions become 
independent of the initial state and thus equal to the random state average, e.g.
\begin{equation}
C(t,t+\tau) = C(\tau) := \aave{V V_\tau}, \quad {\rm if} \quad t \gg t_{\rm E}.
\end{equation}
In this situation we can safely assume random initial states, or argue that for times
larger than $t_E$ the results are the same for any inital state. Then the linear response 
formula can be rewritten as
\begin{equation}
\aave{F(t)}=1-\frac{2\delta^2}{\hbar^2}\left\{\frac{1}{2}t C(0) + \sum_{t'=1}^{t-1}{(t-t')C(t')} 
\right\} + {\cal O}(\delta^3).
\label{eq:FC}
\end{equation}
Now we shall assume that correlation function $C(t)$ decays sufficiently fast, {\em i.e.} faster than 
${\cal O}(t^{-1})$ and that a certain effective time-scale $t_{\rm mix}$ of decay of $C(t)$ exist.
For times $t \gg t_{\rm mix}$ we can neglect the second term under the summation in (\ref{eq:FC}) 
and obtain a linear decay in time $t$ in the linear response regime
\begin{equation}
F_{\rm em}(t) = 1 - 2(\delta/\hbar)^2 \sigma t
\qquad{\rm with}\qquad
\sigma = \frac{1}{2}C(0) + \sum_{t=1}^\infty{C(t)}.
\label{eq:sigma}
\end{equation}
Here, $\sigma$ plays the role of a transport coefficient.

We can make a stronger statement valid beyond the linear response regime if we make an additional assumption 
on the factorization of higher order time-correlations, namely that of $n-$point mixing. This implies that 
$2m$-point correlation $\aave{V_{t_1}\cdots V_{t_{2m}}}$ is appreciably different from zero for 
$t_{2m}-t_1 \to \infty$ only if all the 
(ordered) time indices $\{t_j,j=1\ldots 2m\}$ are {\em paired} with the time differences 
within each pair $t_{2j}-t_{2j-1}$ being of the order or less than $t_{\rm mix}$. 
Here we have assumed without loss of generality that the perturbation is traceless
$\aave{V} = 0$. If not, we subtract $\aave{V}\one$ from $V$ which does not affect probability $F(t)$.
Then we can make a further reduction, namely if $t\gg m t_{\rm mix}$ 
\begin{equation}
\!\op{\cal T}\!\!\!\!\!\sum_{t_1,\ldots,t_{2m}=0}^{t-1}\!\!\!\!
\aave{V_{t_1} V_{t_2} \cdots V_{t_{2m}}}
\rightarrow
\op{\cal T}\!\!\!\!\!\sum_{t_1,\ldots,t_{2m}=0}^{t-1}\!\!\!\!
\aave{V_{t_1} V_{t_2}}\cdots\aave{V_{t_{2m-1}} V_{t_{2m}}}
\rightarrow
\frac{(2m)!}{m!}(t\sigma)^m. \!\!\!\!\!\!\!\!\!\!\!
\label{eq:factor}
\end{equation}
We then obtain a global exponential decay 
\begin{equation}
F_{\rm em}(t)=\left|\aave{M_\delta(t)}\right|^2 = \exp{(-2t/\tau_{\rm em})}, 
\qquad \qquad \tau_{\rm em}=\frac{\hbar^2}{\delta^2 \sigma},
\label{eq:Fmix}
\end{equation}
with a time-scale $\tau_{\rm em}={\cal O}(\delta^{-2})$. 
We should stress that the above result (\ref{eq:Fmix}) has been derived 
under the assumption of
true quantum mixing \cite{qmix}, 
which can be justified only in the limit ${\cal N}\rightarrow\infty$, e.g.
either in semiclassical or thermodynamic limit where the correction (\ref{eq:1N}) is irrelevant.
Note that the result (\ref{eq:Fmix}) can be connected to Fermi golden rule \cite{Beenakker} as
it is based on time-dependent perturbation theory.

\subsection{Non-mixing and non-ergodic regime}

The opposite situation of non-mixing and non-ergodic quantum dynamics, which typically corresponds 
to integrable, near-integrable (KAM), or mixed classical dynamics, is characterized by a non-vanishing 
time-average of the correlation function
\begin{equation}
\bar{C}=\lim_{t\to \infty}{\frac{1}{t^2}\sum_{t',t''=0}^{t-1} C(t',t'')}.
\label{eq:Cinfty}
\end{equation}
Here, due to non-ergodicity, the time-average $\bar{C}$ depends on the structure of the initial state 
$\ket{\psi}$. We further assume that a certain characteristic averaging time-scale $t_{\rm ave}$ exists, 
namely it is an effective time $t=t_{\rm ave}$ at which the limiting process 
(\ref{eq:Cinfty}) converges.
Therefore, for sufficiently large times $t \gg t_{\rm ave}$, the double sum on RHS of
eq. (\ref{eq:F2nd}) can be approximated as $\bar{C}t^2$, so the linear-response formula (\ref{eq:F2nd}) 
yields, in contrast to (\ref{eq:sigma}), a {\em quadratic decay} in time
\begin{equation}
F_{\rm ne}(t)=
1 - \left( \frac{t}{\tau_{\rm ne}}\right)^2 + {\cal O}(\delta^3), \qquad
\tau_{\rm ne}=\frac{\hbar}{\delta \sqrt{\bar{C}}},
\label{eq:Fr2}
\end{equation}
with time-scale $\tau_{\rm ne}={\cal O}(\delta^{-1})$. 
One should observe that the non-ergodic time-scale 
$\tau_{\rm ne}$ can be much smaller than the ergodic-mixing time-scale $\tau_{\rm em}$ (\ref{eq:Fmix})
provided $\hbar$ is fixed, or the limit $\delta\to 0$ is taken prior to the limit $\hbar\to 0$.
Yet $\tau_{\rm ne}$ is typically still much longer than the Zeno time scale of universal
quadratic decay.

Again we can make a much stronger general statement going beyond the second order
$\delta$-expansion. If we assume that $t\gg t_{\rm ave}$, we can re-write the $m$-tuple sums in the 
series (\ref{eq:Msum}) in terms of a {\em time average perturbation operator}
\begin{equation}
\bar{V}=\lim_{t \to \infty}{(1/t)\sum_{t'=0}^{t-1}{V_{t'}}},
\end{equation}
namely
\begin{equation}
f_{\rm ne}(t)=\sum_{m=0}^\infty {\frac{(i \delta t)^m}{\hbar^m m!} \ave{\bar{V}^m}}
=\ave{\exp{(i t \bar{V} \delta/\hbar)}},
\quad
F_{\rm ne}(t) = |f_{\rm ne}(t)|^2.
\label{eq:Favg}
\end{equation}  
Note that $\bar{V}$ is by construction an {\em integral of motion} \cite{PeresInt}, 
$[U,\bar{V}]\equiv 0$, and reduces to a trivial multiple of identity in the case of ergodic
dynamics studied in previous subsection. Whereas in an ergodic and mixing case, 
$m-$th order term of (\ref{eq:Msum}) grows with time only as 
${\cal O}(t^{m/2})$ (for even $m$) since it is dominated by pair time correlations, here 
in a non-ergodic case, the non-trivial time average operator 
$\bar{V}$ already gives the 
dominant effect, namely ${\cal O}(t^m)$ for $m-$th order term of (\ref{eq:Msum}), so the effect of pair 
time correlations can safely be neglected for sufficiently long times ($t\gg t_{\rm ave}$).
Observe also that time averaged correlation is just a variation of the time averaged perturbation
\begin{equation}
\bar{C} = \ave{\bar{V}^2} - \ave{\bar{V}}^2.
\end{equation}
In the semiclassical regime for random initial states, 
$\bar{C}$ goes to a purely classical ($\hbar$-independent) quantity 
$\bar{C}_{\rm cl} = \ave{\bar{v}^2}_{\rm cl} - \ave{\bar{v}}^2_{\rm cl}$
where $\bar{v}$ is a time-averaged classical limit of observable $V$ and $\ave{\bullet}_{\rm cl}$ is
a classical (microcanonical) phase-space average.

Our conclusions may not be valid in the special case where the time-average $\bar{V}$ is
a trivial operator, namely when it either vanishes or is proportional to identity. 
This can happen for 
very special choices of perturbations $V$ or for systems and perturbations with particular geometric or algebraic
symmetries. This option also implies that the classical time-average is trivial $\bar{v}\equiv {\rm const}$, and 
that $\bar{C}=0$.
Of course, in such a case, fidelity decay has to be discussed separately, see e.g. Refs.\cite{newpz,bruno}. 

Let us now use expression (\ref{eq:Favg}) to derive some explicit semiclassical
results in the special case of integrable classical dynamics. For a system with $d$ degrees of
freedom we thus have $d$ canonical constants of motion -- the action variables 
$\ve{I}=(I_1,\ldots,I_d)$, which
are quantized using EBK rule $\ve{I}_\ve{n}=\hbar(\ve{n}+\ve{\gamma}/4)$ where 
$\ve{n}=(n_1,\ldots,n_d)$ is a vector of integer quanum numbers $n_j$ and 
$\ve{\gamma}=(\gamma_1,\ldots,\gamma_d)$ is a vector of integer Maslov 
indices $\gamma_j$;
the latter are irrelevant for the discussion that follows. 
Since the time averaged operator $\bar{V}$ commutes with $U$ and with 
the actions $\ve{I}$, it is diagonal in the (generically non-degenerate) basis of
eigenstates of $\ve{I}$ (quantized tori) $\ket{\ve{n}}$. In leading semiclassical order
one may write
\begin{equation}
\bra{\ve{n}}\bar{V}\ket{\ve{n}'} = 
\delta_{\ve{n},\ve{n}'}\bar{v}(\ve{I}_\ve{n})
\end{equation}
where $\bar{v}(\ve{I})$ is the corresponding classical time-averaged observable in action space.
The fidelity (\ref{eq:Favg}) can therefore be written as
\begin{equation}
f_{\rm ne}(t) = \sum_{\ve{n}}\exp(i t \bar{v}(\ve{I}_\ve{n}) \delta/\hbar)|\braket{\ve{n}}{\psi}|^2.
\label{eq:Fscs}
\end{equation}
Provided the diagonal elements of the density matrix can be written in terms of some smooth 
{\em structure function} $D(\ve{I}_{\ve{n}}) = |\braket{\ve{n}}{\psi}|^2$, 
and replacing the sum (\ref{eq:Fscs}) by an integral over the action space, which is justified for 
small $\hbar$ up to classically long time $\propto \hbar^0 \delta^{-1}$, we obtain
\begin{equation}
f_{\rm ne}(t)=\hbar^{-d} \int\! d^d\ve{I}\, \exp{\{i t \bar{v}(\ve{I})\delta/\hbar\}}D(\ve{I}). 
\label{eq:Fsemi}
\end{equation}
The obvious next step is to compute this integral by a method of stationary phase.
However the result depends on the precise form of the function $D(\ve{I})$ 
which may in turn depend explicitly 
on $\hbar$. Below we work out the details for two important special cases, 
namely a random
and a coherent initial state.

\subsubsection{Semiclassical asymptotics for a random initial state.}

Let us first assume uniform averaging over (random) initial states
$D(\ve{I}) \equiv 1/{\cal N} = (2\pi\hbar)^d/{\cal V}$. For large $t \delta/\hbar$ the above 
integral (\ref{eq:Fsemi}) can be written as a sum of contributions stemming from, say $p$ points, 
$\ve{I}_\eta, \eta=1,\ldots,p$ where the phase is stationary, 
$\partial\bar{v}(\ve{I}_\eta)/\partial\ve{I}=0$. This yields
\begin{equation}
f_{\rm ne}^{\rm ave}(t) = \frac{(2\pi)^{3d/2}}{\cal V}\left|\frac{\hbar}{t\delta}\right|^{d/2}\sum_{\eta=1}^p
\frac{\exp\{i t\bar{v}(\ve{I}_\eta)\delta/\hbar + i \nu_\eta \}}{|\det \ma{\bar{V}}_\eta|^{1/2}},
\label{eq:Fsqrt}
\end{equation}
where 
$\{\ma{\bar{V}}_\eta\}_{jk} := \partial^2 \bar{v}(\ve{I}_\eta)/\partial I_j\partial I_k$ is a matrix of second
derivatives at the stationary point $\eta$, and $\nu_\eta = \pi(m_+ - m_-)/4$
where $m_{\pm}$ are the numbers of positive/negative eigenvalues of the matrix
$\ma{\bar{V}}_\eta$. 
The stationary phase formula (\ref{eq:Fsqrt}) is expected to be correct in the range 
${\rm const}\,\hbar/\delta < t < {\rm const}'/\delta$.
Most interesting to note is the asymptotic power-law time and perturbation dependence 
$F^{\rm ave}_{\rm ne} \sim |\hbar/(t\delta)|^d$, which allows for a possible crossover to a 
Gaussian decay\cite{Prosen01} when approaching the thermodynamic limit $d\to\infty$.

\subsubsection{Semiclassical asymptotics for a coherent initial state.}

Now let us consider a single $d$-dimensional general coherent state centered at 
$(\ve{I}^*,\ve{\theta}^*)$ in action-angle space
\begin{equation}
\braket{\ve{n}}{{\ve{I}^*,\ve{\theta}^*}} = 
\left(\frac{\hbar}{\pi}\right)^{d/4}
\!\!\!\left|\det\Lambda\right|^{1/4}
\exp\left\{-\frac{1}{2\hbar}(\ve{I}_{\ve{n}} - \ve{I}^*)\cdot\Lambda(\ve{I}_{\ve{n}}-\ve{I}^*) - 
i\ve{n}\cdot\ve{\theta}^*\right\},
\label{eq:CS}
\end{equation}
where $\Lambda$ is a positive symmetric $d\times d$ matrix of squeezing 
parameters, giving 
\begin{equation}
D(\ve{I}) = (\hbar/\pi)^{d/2}\left|\det\Lambda\right|^{1/2}
\exp(-(\ve{I}-\ve{I}^*)\cdot\Lambda (\ve{I}-\ve{I}^*)/\hbar)
\end{equation}
and
\begin{equation}
f^{\rm coh}_{\rm ne}(t) = \frac{\left|\det\Lambda\right|^{1/2}}{(\pi\hbar)^{d/2}}
\int\!d^d\ve{I}\,\exp\left\{\frac{1}{\hbar}
\left(i t \bar{v}(\ve{I})\delta 
- (\ve{I}-\ve{I}^*)\cdot\Lambda(\ve{I}-\ve{I}^*)
\right)\right\}.
\label{eq:Fcohst}
\end{equation}
Using the assumption $\delta t\ll 1$, we see that a unique stationary point $\ve{I}_s$ of 
the exponent approaches $\ve{I}^*$ as $\delta\to 0$, 
\begin{equation}
\ve{I}_s = \ve{I}^* - \frac{it\delta}{2}\Lambda^{-1}\ve{v}' + {\cal O}(\delta^{2}),
\quad {\rm where}\quad
\ve{v}' := \frac{\partial\bar{v}(\ve{I}^*)}{\partial\ve{I}} 
\end{equation}
so we may explicitly evaluate (\ref{eq:Fcohst}) by the method of stationary phase without any 
lower bound on the range of time $t$,
\begin{equation}
f^{\rm coh}_{\rm ne}(t) = \exp\left\{-\frac{(\ve{v}'\cdot \Lambda^{-1}\ve{v}')\delta^2}{4\hbar}t^2 + 
\frac{i\bar{v}(\ve{I}^*)\delta}{\hbar} t\right\}.
\label{eq:Fcoh}
\end{equation}
Note that the fidelity decay for a coherent initial state with regular classical motion has a 
time-scale
\begin{equation}
\tau_{\rm ne-coh} = (2\hbar)^{1/2}(\ve{v}'\cdot\Lambda^{-1}\ve{v}')^{-1/2}\delta^{-1}
\;\;\propto\;\; \hbar^{1/2}\delta^{-1},
\label{eq:taunecoh}
\end{equation}
which is consistent with (\ref{eq:Fr2}) with 
$\bar{C}=\frac{1}{2}\hbar(\ve{v}'\cdot\Lambda^{-1}\ve{v}')$
and is by a factor proportional to $\hbar^{-1/2}$
longer than the time-scale 
$\tau_{\rm ne}\propto \hbar/\delta$ for a random initial state.
It should be noted that the above derivation of fidelity decay for a coherent initial state 
(\ref{eq:CS}-\ref{eq:taunecoh}) remains valid in a near-integrable (KAM) situation 
of mixed classical phase space, provided that the initial wave packet is launched in a 
{\em regular region} of phase space where (local) action-angle variables exist.

\subsection{Finite size effects and time and perturbation scales}   
\label{sec:timepert}

The theoretical relations of the previous subsections are strictly justified in the 
asymptotic limit ${\cal N}\to\infty$. For finite ${\cal N}$, fidelity 
$F(t)$ cannot decay indefinitely but starts to fluctuate 
for long times due to discreteness of the spectrum of the evolution operator $U$. 
Let us write the eigenphases of $U$ and $U_\delta$, and the
corresponding eigenvectors, respectively, as $\phi_n$, $\phi^\delta_n$, and 
$\ket{\phi_n}$, $\ket{\phi^\delta_n}$, $n=1,\ldots,{\cal N}$, 
satisfying $U\ket{\phi_n}=e^{-i\phi_n}\ket{\phi_n}$,
$U_\delta\ket{\phi^\delta_n}=e^{-i\phi^\delta_n}\ket{\phi^\delta_n}$.
Now define a unitary operator $W$ which maps the eigenbasis of 
$U$ to the eigenbasis of $U_\delta$, namely $W\ket{\phi_n}:=\ket{\phi^\delta_n}$ for all $n$, with 
matrix elements
$W_{mn} := \bra{\phi_m}W\ket{\phi_n}$, and write the initial state as
$\psi_n = \braket{\phi_n}{\psi}$.
Note that the matrix $W_{mn}$ is {\rm real orthogonal} if $U$ and $U_\delta$ possess a common 
anti-unitary symmetry (e.g. time-reversal).
Now it is straightforward to rewrite the fidelity 
amplitude as
\begin{equation}
f(t) = \sum_{l,m,n} \psi^*_n \psi_m W^*_{nl} W_{ml} 
\exp\left(i(\phi_n-\phi^\delta_l)t\right),
\label{eq:Ftfin}
\end{equation}
At this point we are interested in the long time fluctuations so we compute 
the time averaged fidelity fluctuation
\begin{equation}
F_{\rm ta} := \lim_{T\to\infty} \frac{1}{T} \int_0^T dt |f(t)|^2 
= \sum_{k,l,m,n} \psi_k^* W_{kl}^* \psi_m W_{m l} |\psi_n|^2 |W_{n l}|^2. 
\label{eq:Frms}
\end{equation}
In the process of averaging over the time we have assumed that the 
eigenphases are non-degenerate so
$\overline{\exp(i(\phi_n-\phi_{n'}+\phi^\delta_l-\phi^\delta_{l'})t)} = \delta_{n n'}\delta_{l,l'}$.
We see that fidelity fluctuation $F_{\rm ta}$ depends on the orthogonal/unitary matrix $W_{mn}$ 
and the initial state. Detailed discussion of various cases are given in Ref.\cite{ktop} Here
we only give results for random intitial state, where $\psi_n$ can be assumed,
for large ${\cal N}$, to be independent complex random Gaussian variables
with variance $1/{\cal N}$. Averaging over an ensemble of initial states 
yields
\begin{equation}
\aave{F_{\rm ta}} = 
\frac{1}{{\cal N}^2}\left(\sum_{n,l} |W_{nl}|^4 + \sum_{n,m,l}|W_{nl}|^2 |W_{ml}|^2\right).
\label{eq:Frmsb}
\end{equation}
For sufficiently weak perturbation, one may assume that the matrix $W_{n,m}$ is
near identity $W_{n,m} \approx \delta_{n,m}$, thus
$F^{\rm weak}_{\rm ta-ii} = 2/{\cal N}$
while for strong perturbation $W_{n,m}$ can be assumed to be a random matrix,
yielding $F^{\rm strong}_{\rm ta-ii}= 1/{\cal N}$.
Therefore the $1/{\cal N}$ dependence of the residual fidelity $F_{\rm ta}$,
for random intial states {\em does not depend on the structure of 
eigenstates} $W_{n,m}$, and there is only a factor of $2$ difference between 
the two extreme cases.

In addition to a random initial state, we shall now assume ergodic and
mixing classical dynamics, so fidelity decay is initially given
by exponential law (\ref{eq:Fmix}) until it reaches the plateau (\ref{eq:Frmsb}). 
The {\em saturation time} $t_*$ where $F_{\rm em}(t_*) = 
F_{\rm ta}$, is the first important time scale
\begin{equation}
t_* = \frac{1}{2}\tau_{\rm em}\ln{\cal N} \approx \frac{\hbar^2 d}{\delta^2\sigma_{\rm cl}}\ln(1/\hbar),
\end{equation}
where $\sigma_{\rm cl}$ is the classical limit of the transport 
coefficient (\ref{eq:sigma}).

The second new time-scale is related to the asymptotic non-decay of time 
correlations for finite-${\cal N}$ quantum dynamics, namely even if the 
system is classically mixing the quantum correlation function will have 
a small non-vanishing ($\hbar$-dependent) time average
\begin{equation}
\bar{C} = \overline{\ave{V(t)V(t')}} = \sum_n |\psi_n|^2 (V_{nn})^2,
\label{eq:diagonal}
\end{equation}
where $V_{mn} = \bra{\phi_m}V\ket{\phi_n}$. Since, however, the 
classical system is assumed to be chaotic, implying ergodicity and mixing,
the matrix elements $V_{mn}$ behave like pseudo-random variables with
a variance given by the Fourier transformation $S(\omega)$ of the 
corresponding classical correlation function $C_{\rm cl}(t)$ at 
frequency $\omega=\phi_m-\phi_n$ \cite{FeingoldPeres}. 
On the diagonal we have $\omega=0$ and an additional factor of $2$
due to the form of the invariant random matrix measure (see e.g. \cite{Haake1}). 
Thus we have
\begin{equation}
\bar{C} = \frac{4\sigma_{\rm cl}}{\cal N},
\label{eq:finitesize}
\end{equation}
where $\sigma_{\rm cl} = S(0)/2$ is the classical limit of (\ref{eq:sigma}).
The result, due to ergodicity, does not depend on the particular form of the
initial state $\ket{\psi}$.
The decay of fidelity (\ref{eq:FC}) will start to be dominated by the 
average plateau (\ref{eq:finitesize})
at sufficiently long time $t$, when
$\sum_{t'=0}^{t-1}{(t-t')\bar{C}} \approx 2\sigma_{\rm cl} t^2/{\cal N} \ge \sigma_{\rm cl} t$,
i.e. for times $t$ greater than $t_{\rm p}$
\begin{equation}
t_{\rm p}=\frac{1}{2}{\cal N} \propto \hbar^{-d},
\label{eq:tp}
\end{equation}
which is just the {\em Heisenberg time} associated to the inverse 
density of states.

\begin{figure}
\centerline{\includegraphics[width=2.5in]{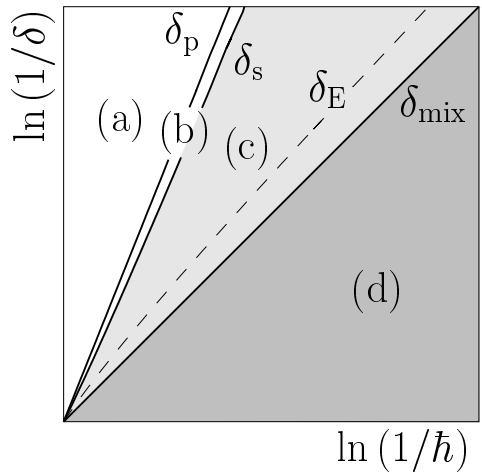}}
\caption{Schematic view of different regimes (a-d) of 
fidelity decay in the regime of classically chaotic (or mixing) dynamics.}
\label{fig:delta}
\end{figure}

Depending on the interrelation among four (or five) time-scales 
$\tau_{\rm em} \propto \hbar^2 \delta^{-2}$,
$t_{\rm p} \propto \hbar^{-d} \delta^0$,
$t_* \propto \hbar^2 \ln(1/\hbar) \delta^{-2} d$,
$t_{\rm mix} \propto \hbar^0 \delta^0$,
(and $t_{\rm E} \propto \ln(1/\hbar)\delta ^0$ if we are considering coherent 
initial states, like e.g. \cite{Jalabert,Pastawski,Beenakker})
we can have four (or five) different regimes depending on the three main scaling parameters: 
perturbation strength $\delta$, Planck's constant $\hbar$, and dimensionality $d$. 
Note that we always have $t_* > \tau_{\rm em}$. All regimes can be reached by changing only 
the parameter $\delta$ while keeping $\hbar$ and $d$ fixed (see fig.~\ref{fig:delta}):
\\\\
({\bf a}) For sufficiently small perturbation $\delta$ we will 
have $t_{\rm p} < \tau_{\rm em}$. 
This means that $F_{\rm em}(t_{\rm p})$ is still close to $1$ and we will 
initially have quadratic decay 
(\ref{eq:Fr2}) with $\bar{C}$ given by an average finite size plateau (\ref{eq:finitesize}). 
This will occur for $\delta \le \delta_{\rm p}$ where
\begin{equation}
\delta_{\rm p}= \hbar \left(\frac{2}{\sigma_{\rm cl}{\cal N}}\right)^{1/2} =
\frac{\sqrt{2}(2\pi)^{d/2}}{({\cal V}\sigma_{\rm cl})^{1/2}}\hbar^{d/2+1}.
\label{eq:deltap}
\end{equation}
In fact, in this regime, also referred to \cite{Beenakker,Tomsovic} as {\em perturbative}, 
one may use first order stationary perturbation theory on the eigenstates of $U_\delta$, yielding
$\phi^\delta_n = \phi_n + V_{nn}\delta/\hbar + {\cal O}(\delta^2)$,
$V_{mn} = \delta_{mn} + {\cal O}(\delta)$, and rewrite (following \cite{Tomsovic}) the
finite size fidelity (\ref{eq:Ftfin}) in terms of a Fourier transform of a probability 
distribution $w(V)$ of 
diagonal matrix elements $V_{nn}$, $f_{\rm pert}(t) = \int dV w(V) \exp(-iVt\delta/\hbar)$.
Since $w(V)$ is conjectured to be Gaussian for classically ergodic and mixing system \cite{FeingoldPeres}, 
it follows that $f_{\rm pert}(t)$ is also a Gaussian with a semiclassically long time-scale $\tau_{\rm p}$
\begin{equation}
f_{\rm pert}(t) = \exp{(-(t/\tau_{\rm p})^2/2)},\quad
\tau_{\rm p} =
\left(\frac{\cal N}{\sigma_{\rm cl}}\right)^{1/2}\frac{\hbar}{2\delta} =
\frac{{\cal V}^{1/2}}{(2\pi)^{d/2}\sigma_{\rm cl}^{1/2}}\frac{\hbar^{1-d/2}}{2\delta}.
\label{eq:Fp} 
\end{equation}
({\bf b}) If $\tau_{\rm em} < t_{\rm p} < t_*$ we will have a crossover from initial exponential decay of fidelity (\ref{eq:Fmix})
to a Gaussian decay (\ref{eq:Fp}) at $t\sim t_{\rm p}$, which will terminate and go over to fluctuating behavior
when $F_{\rm pert}(t) = F_{\rm ta}({\cal N})$. Note that this will happen before time 
$t_*$ which is estimated based on a slower exponential decay (\ref{eq:Fmix}).
This regime will exist in perturbation range $\delta_{\rm p} < \delta < \delta_{\rm s}$ with an upper border $\delta_{\rm s}$ 
determined by the condition $t_{\rm p}=t_*$ to be
\begin{equation}
\delta_{\rm s}=(\ln{\cal N}/2)^{1/2} \delta_{\rm p}.
\label{eq:deltas}
\end{equation} 
({\bf c}) If we still further increase $\delta$, we have the most 
interesting, `fully nonperturbative' regime,
when $\tau_{\rm em} < t_* < t_{\rm p}$ and we will have a full exponential decay (\ref{eq:Fmix}), up to time $t_*$ 
when the fidelity reaches finite size fluctuations.
This regime continues as long as $\delta < \delta_{\rm mix}$ where the border
\begin{equation}
\delta_{\rm mix} = \frac{\hbar}{\sqrt{\sigma_{\rm cl}t_{\rm mix}}} = 
\left(\frac{\cal N}{2t_{\rm mix}}\right)^{1/2} \delta_{\rm p}
\label{eq:deltam}
\end{equation}
is determined by the condition $\tau_{\rm em} = t_{\rm mix}$ which is a point
where the arguments leading to the factorization (\ref{eq:factor})
and exponential decay (\ref{eq:Fmix}) are no longer valid.
We note that the relative size of this window range
$\delta_{\rm mix}/\delta_{\rm s}=\sqrt{{\cal N}/(t_{\rm mix}\ln{\cal N})}$ increases, both, 
in the semiclassical and in the thermodynamic limit. 
This regime also corresponds to `Fermi golden rule decay'
discussed in \cite{Beenakker}.
\\\\ 
({\bf d}) Further increasing $\delta > \delta_{\rm mix}$, the estimated 
fidelity decay time eventually becomes smaller than the classical mixing time 
$t_{\rm mix}$, $\tau_{\rm em} < t_{\rm mix}$. In this regime, the perturbation
is simply so strong that the fidelity
effectively decays within the shortest observable time-scale ($t_{\rm mix}$).

However if we consider a non-random, e.g. {\em coherent initial state} then
the quantum correlation function relaxes on a slightly longer, namely 
Eherenfest time-scale $t_{\rm E}$ so the regime (c) should terminate 
already at a little smaller upper border $\delta=\delta_{\rm E}$ which 
is naturally determined by 
the condition $\tau_{\rm em} = t_{\rm E}$
\begin{equation}
\delta_{\rm E} \approx 
\hbar\frac{\lambda^{1/2}}{[\sigma_{\rm cl}\ln(1/\hbar)]^{1/2}} 
\sim \frac{\delta_{\rm mix}}{[\ln(1/\hbar)]^{1/2}}.
\end{equation}
For coherent initial states one thus obtains an extra but very narrow regime 
$\delta_{\rm E} < \delta < \delta_{\rm mix}$ (describing the
time-range $t_{\rm mix} < t < t_{\rm E}$) where the fidelity decay can be
computed in terms of classical Lyapunov exponents \cite{Jalabert,Pastawski}.

Yet in the regime of non-ergodic, say integrable classical
mechanics things are simpler, as we do not have to worry about the 
average plateau in the correlation function due to a 
finite ${\cal N}$ because we already have a higher average time 
correlation $\bar{C}\to \bar{C}_{\rm cl}\neq 0$ (\ref{eq:Cinfty}).
Thus we have here only two relevant 
time-scales, namely $\tau_{\rm ne}$ giving initial quadratic decay 
(\ref{eq:Fr2}), and the saturation time-scale $t_*$, due to finite size 
fidelity fluctuation, which depends on the properties of the
initial state (power law (\ref{eq:Fsqrt}) for a random initial state, 
versus Gaussian (\ref{eq:Fcoh}) for a coherent initial state).
However in the particular special case of `ergodic' perturbation $\bar{V}=0$ 
of regular dynamics (as mentioned above), one now clearly
obtains that $\bar{C}=0$ which is equivalent of having 
all diagonal matrix elements vanishing, {\it i.e.} $V_{nn}=0$ [see eq. (\ref{eq:diagonal})]. 
In this case the time scale of fidelity decay $\tau_{\rm ne}$ from eq. (\ref{eq:Fr2}) formally 
diverges. This in fact means that fidelity decays much slower than in generic case, now the
decay being described by the first nonvanishing -- {\em i.e.} fourth order term in the 
$\delta$-expansion.\cite{newpz}

Note that our result is not contradicting any of the known facts of the
quantum-classical correspondence. For example, a growth of quantum 
dynamical entropies \cite{Alicki} persists only up to logarithmically 
short Ehrenfest time $t_{\rm E}$, which is the upper bound for the 
validity of the strict classical-quantum correspondence in the fidelity decay
as derived in Ref.\cite{Jalabert,Pastawski} and within which one would always find 
$F^{\rm coh}_{\rm ne}(t) > F_{\rm em}(t)$ above the perturbative 
border $\delta > \delta_{\rm p}$, whereas our theory reveals new 
nontrivial quantum phenomena with a semiclassical prediction (but 
not correspondence!) much beyond that time. On one hand, if we let $\hbar\to 0$ 
first, and then $\delta \to 0$, we recover a result supported 
by a classical intuition, namely that the regular (non-ergodic) dynamics
is more stable than the chaotic (ergodic and mixing) dynamics. 
On the other hand, if we let $\delta\to 0$ first, and only after that $\hbar\to 0$,
we find somewhat counterintuitive results saying that chaotic (mixing) 
dynamics is more stable than the regular one.

\section{Purity fidelity}
\label{sec:pf}

In this section we propose another characteristic which describes
the quality of Loschmidt echoes, in particular for systems
composed of two parts corresponding to different degrees of freedom, 
for example the central system and the environment.

We consider a system with a product Hilbert space
${\cal H}={\cal H}_1\otimes {\cal H}_2$,
where ${\cal H}_{1,2}$ are factor Hilbert spaces of the two subsystems.
We are again studying an echo experiment, now with the the initial state 
beeing a product state
\begin{equation}
\ket{\psi(0)} = \ket{\psi_1}\otimes \ket{\psi_2},
\end{equation}
yielding a, generally {\em entangled}, final state
\begin{equation}
\ket{\psi(t)} = M_\delta(t)\ket{\psi(0)}.
\end{equation}
The entanglement produced by imperfect quantum echo can be most directly
characterized by the {\em reduced density matrix}
\begin{equation}
\rho_1(t) = \tr_2 \ket{\psi(t)}\bra{\psi(t)} = 
\tr_2 \left( M_\delta(t) \ket{\psi(0)}\bra{\psi(0)} M^\dagger_\delta(t) \right)
\label{eq:rho1}
\end{equation}
obtained by tracing out the second subsytem ${\cal H}_2$.
Now one can say that $\ket{\psi(t)}$ is disentangled iff the
density matrix $\rho_1(t)$ is pure, and entanglement can be characterized by
purity; specifically for the echo situation 
we call this quantity {\em purity fidelity}
\begin{equation}
F_{\rm P}(t) = \tr_1 [\rho_1(t)]^2.
\label{eq:purfid}
\end{equation}
Note that purity fidelity is in some sense a weaker quantity than fidelity:
in order for fidelity to be high, purity fidelity must also be high, but
not vice versa.
High purity fidelity only requires the state to return to the factorized form,
which is a much weaker condition than complete recurrence as required by high fidelity.

Now we apply our perturbative expansion of the echo operator (\ref{eq:Msum}) 
to the expression for purity fidelity (\ref{eq:purfid}). Keeping only terms 
up to second order in the perturbation strength $\delta$ we obtain a linear 
response formula for purity fidelity, 
\begin{eqnarray}
F_{\rm P}(t) &=& 1 - 
2 \delta^2 \hbar^{-2} \left\{C_{\rm int}(t) - D_{\rm int}(t)\right\} +\cdots, \label{eq:Fp2nd} \\
D_{\rm int}(t)&:=& \sum_{\nu \neq 1} |\bracket{1,\nu}{\Sigma(t)}{1,1}|^2 + 
\sum_{i \neq 1}{|\bracket{i,1}{\Sigma(t)}{1,1}|^2}, 
\nonumber
\end{eqnarray}
which is analogous to a simpler formula (\ref{eq:F2ndb}) for fidelity.
We have used an explicit notation of a complete basis of a product Hilbert space as
$\ket{i,\nu}=\ket{i}\otimes\ket{\nu}$ where latin indeces label states in the 
first and greek indeces states in the second subspace.
There is a general relation between fidelity and purity fidelity which can be formulated in terms of a 
rigorous mathematical inequality\cite{ineq}
\begin{equation}
F_{\rm P}(t) \ge F^2(t).
\label{eq:ineq}
\end{equation}

The formal similarity of linear response formulae (\ref{eq:F2ndb}) and
(\ref{eq:Fp2nd}) results in similar physical behaviour of fidelity and purity fidelity,
in both qualitatively different cases of dynamics, ergodic and mixing --- chaotic, and
regular.
For example, in the regime of ergodic and mixing dynamics we can use the
result that fidelity is after a while, due to ergodicity, independent of the
initial state, hence we can write for the echo operator in the weak limit sense
\begin{equation}
M_\delta(t) \to \exp(-t/\tau_{\rm em}) \one.
\end{equation}
Plugging this expression into formula (\ref{eq:rho1}) and the obtained result into 
formula (\ref{eq:purfid}) we finally obtain a simple expression compatible with
the upper bound of the inequality (\ref{eq:ineq})
\begin{equation}
F_{\rm P}(t) = \exp(-4t/\tau_{\rm em}).
\end{equation}
In this case the relative weight of the correction term $D_{\rm int}$ in the
linear response formula (\ref{eq:Fp2nd}) vanishes in proportion to $C_{\rm int}$
in the semiclassical limit $\hbar\to 0$, ${\cal N}\to\infty$.
Again, the same consideration of time and perturbation scales applies for finite 
${\cal N}$ as discussed above for the case of fidelity.

In the non-mixing or classically regular case, the situation
is more complicated as purity fidelity depends on the structure of initial state 
$\ket{\psi}$. The only thing we can state generally in this case is the quadratic 
decay in the linear response regime, i.e. as long as $1-F_{\rm P}(t)$ is small we have
\begin{equation}
F_{\rm P}(t) = 1 - \frac{2\delta^2}{\hbar^2}(\bar{C}-\bar{D})t^2 + {\cal O}(t^3).
\end{equation}
Since in this case time-averaged correlation functions are non-vanishing,
the integrated ones grow as $\propto t^2$, so we have written 
$C_{\rm int}(t) = \bar{C}t^2$, 
$D_{\rm int}(t) = \bar{D}t^2$.
It is interesting to discuss the two extreme cases of intial states: 
(i) In the case of coherent initial states (Gaussian wave packets) we have 
shown \cite{JCPRL} that the term $\bar{D}$ cancels the term $\bar{C}$ in the
leading semiclassical order, so $\bar{C}-\bar{D} \propto \hbar^2$, whereas
$\bar{C} \propto \hbar$. This means that purity fidelity for coherent initial states
decays with an $\hbar$ independent time scale which is by a factor proportional to $\hbar^{-1/2}$
longer than the time scale of fidelity decay.
(ii) On the other hand, in the case of initial random states one can show that the term $\bar{D}$ is 
negligible compared to $\bar{C}$ in the semiclassical limit, so both, fidelity, and purtity fidelity
decay with the same rate, simliarly as in the case of chaotic dynamics.

There is an important special case of
potential practical interest, if the perturbation is such that
the unperturbed composite system is {\em decoupled}, i.e. $U_{\delta=0} = U_1 \otimes U_2$.
Then the unperturbed (e.g. forward) evolution does not change the purity, so the
purity fidelity (purity of echo dynamics) is the same as purity (or linear entropy) of uni-directed 
time-evolution alone. Again, our linear response formalism predicts faster increase of
linear entropy (decay of purity) for regular or weakly chaotic systems than 
for strongly chaotic systems.\cite{ktop2}
This result has been independently reproduced in Ref.\cite{Tanaka}.

In next section we outline some of our numerical results which confirm the theory of the last 
two sections.

\section{Numerical experiments}

We shall  consider two numerical toy models by which we may demonstrate and verify the
theoretical results of previous sections.

\begin{figure}
\centerline{\includegraphics{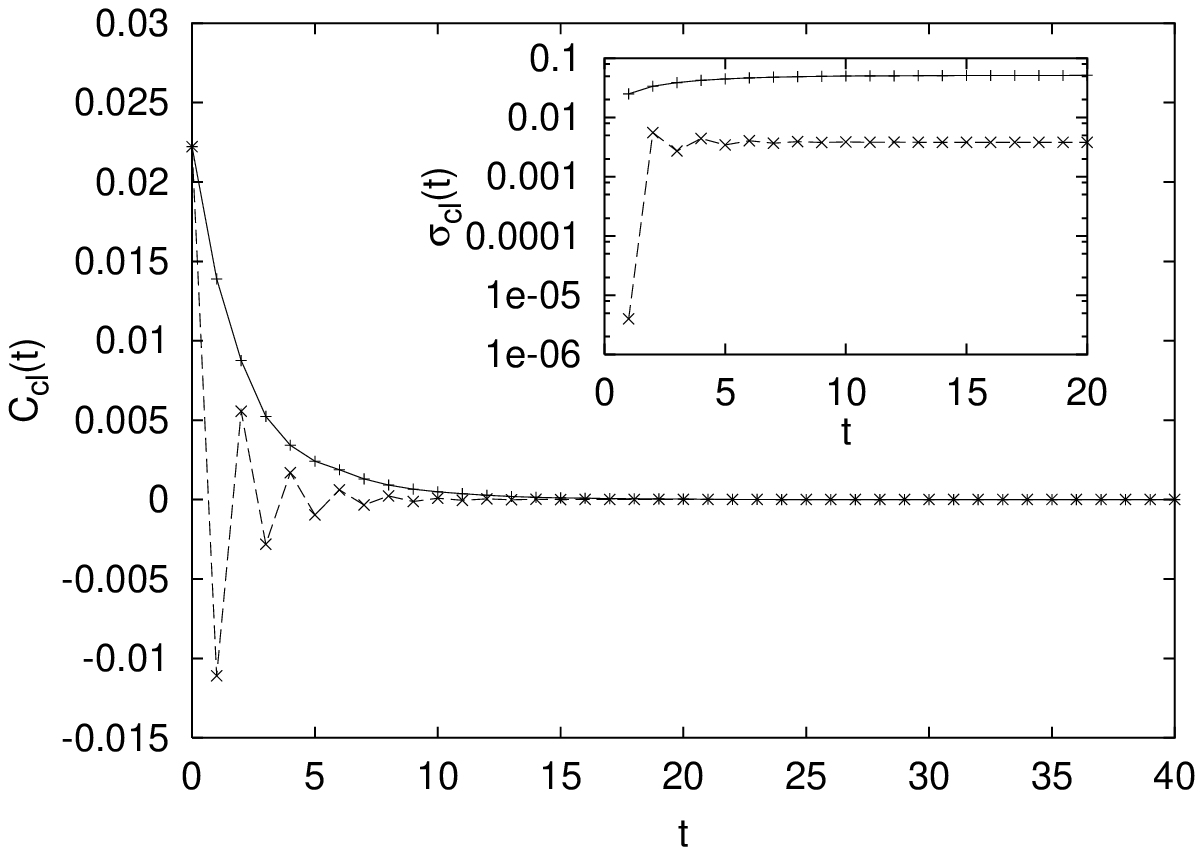}}
\caption{Classical correlation function $C_{\rm cl}(t)$ for kicked top with
$\alpha=30$, and $\gamma=\pi/6$ (top, full curve) and $\gamma=\pi/2$ (bottom, broken curve). Finite time 
integrated correlation function is shown in the inset, converging to $\sigma_{\rm cl}=0.00385$ and 
$0.0515$, for $\gamma=\pi/2$ and $\pi/6$, respectively. Averaging over $10^5$ random initial 
conditions on a sphere is performed.}
\label{fig:class30}
\end{figure}

First we choose Haake's quantized kicked top \cite{Haake2} since this model served as a 
model example for many related studies, see e.g. Refs.\cite{Shack94,Beenakker,Alicki,Fox,Casati,Breslin,Haake00}.
The unitary propagator of the kicked top reads
\begin{equation}
U = U(\alpha,\gamma)=\exp{(- i \gamma J_{\rm y})} \exp{(-i \alpha J_{\rm z}^2/2J)},
\label{eq:Ukt}
\end{equation}
where $J_k$ ($k={\rm x,y,z}$) are quantum angular momentum operators obeying 
$[J_k,J_l]=i \epsilon_{klr} J_r$. The (half)integer $J$ determines the size of the Hilbert space 
$2J+1$ and the value of the effective Planck constant $\hbar=1/J$.
The perturbation is defined by varying the parameter $\alpha$, 
$U_\delta=U(\alpha+\delta,\gamma)$, so $V$ reads
\begin{equation}
V=\frac{1}{2}\left(\frac{J_{\rm z}}{J}\right)^2.
\label{eq:KTa}
\end{equation}
The classical limit is obtained by letting $J = 1/\hbar \to \infty$ and writing the classical angular 
momentum in terms of a unit vector on a sphere $\ve{r} = (x,y,z) = \ve{J}/J$.
The Heisenberg equation for the SU(2) operators $\ve{J}$, $\ve{J}' = U^\dagger \ve{J} U$, reduces to 
the classical area preserving map of a sphere
\begin{eqnarray}
x'&=& \cos{\gamma}(x \cos{\alpha z}-y \sin{\alpha z})+z\sin{\gamma} \nonumber \\
y'&=& y \cos{\alpha z} + x \sin{\alpha z} \nonumber \\
z'&=& z \cos{\gamma}+\sin{\gamma}(y\sin{\alpha z}-x\cos{\alpha z}).
\label{eq:KTclass}
\end{eqnarray}
Note that in the classical limit the perturbation generator is
\begin{equation}
v(\ve{r})=\frac{z^2}{2}.
\label{eq:Aclass}
\end{equation}
For $\alpha=0$ the system is integrable, while with increasing $\alpha$ there is a transition to 
chaotic motion. The second parameter $\gamma$ is usually set to $\pi/2$, however in our numerical 
simulation we will use two different values exhibiting qualitatively different correlation decay 
(for large $\alpha$): the 'standard' case $\alpha=30,\gamma=\pi/2$ where $C_{\rm cl}(t)$ decays in oscillatory 
way and the case $\alpha=30,\gamma=\pi/6$ where $C_{\rm cl}(t)$ decays monotonically
(see fig.~\ref{fig:class30}).
\par
In the case of $\gamma=\pi/2$ we have two discrete symmetries. The evolution $U$ 
commutes with $R_{\rm x}$ and $R_{\rm y}$, the rotations of $\pi$ around 
the ${\rm x}$ and ${\rm y}$ axes, respectively. 
The Hilbert space is therefore reducible into three invariant subspaces (using notation of Peres's book \cite{Peres1} with the
basis $\ket{m}$ of eigenstates of $J_{\rm z}$ and assuming $J$ to be an {\em even integer}): 
EE of dimension $J/2+1$ with the basis states $\ket{0}$ and $\{ \ket{2m}+\ket{-2m} \}/\sqrt{2}$; 
OO of dimension $J/2$ with the basis $\{ \ket{2m-1}-\ket{-(2m-1)} \}/\sqrt{2}$; 
OE of dimension $J$ with the basis $\{ \ket{2m}-\ket{-2m}\}/\sqrt{2}$ and $\{ \ket{2m-1}+\ket{-(2m-1)}\}/\sqrt{2}$ 
with $m=1,\ldots,J/2$ in all three cases. For $\gamma \neq \pi/2$ the spaces OO and EE coalesce as 
$R_{\rm y}$ is the only discrete symmetry left.
In numerical experiments we always choose the OE subspace so that the dimension of the Hilbert space is ${\cal N}=J$. 
\par
We will compute the fidelity of two different types of initial states:
(1) random initial state with components $c_m = \braket{m}{\psi}$ being independent Gaussian 
pseudo-random numbers, and (2) pure minimal wavepacket initial state, 
namely SU(2) coherent state $\ket{\psi}=\ket{\vartheta,\varphi}$ centered at the point 
$\ve{n}=(\sin{\vartheta}\cos{\varphi},\sin{\vartheta}\sin{\varphi},\cos{\vartheta})$ on a unit sphere
\begin{equation}
\ket{\vartheta,\varphi}=\sum_{m=-J}^{J}{\left( {2J \atop J+m} \right)^{1/2} 
\cos{(\vartheta/2)}^{J+m} \sin{(\vartheta/2)}^{J-m} e^{-i m \varphi} \ket{m}}.
\label{eq:SU2coh}
\end{equation}

\begin{figure}
\centerline{\includegraphics[width=110mm]{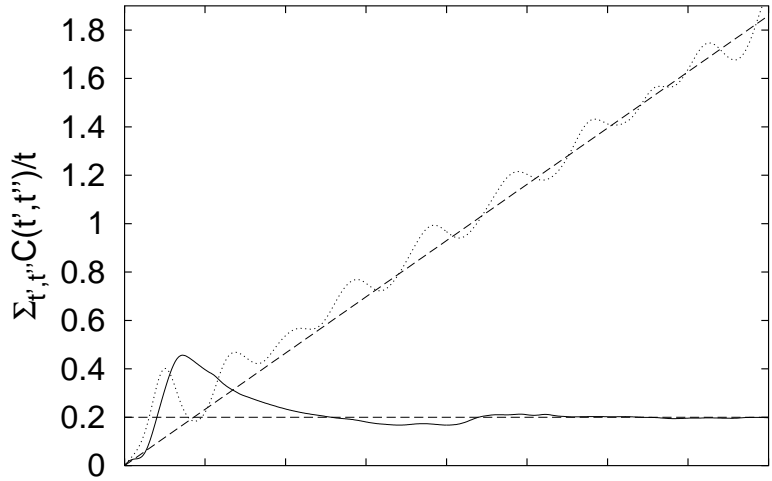}}
\caption{
Integrated quantum correlation function of the perturbation in the JC model, divided
by time $t$. Solid curve corresponds to classically chaotic case (see text) and
is approaching a constant plateau (dashed), whereas dotted curve corresponds to integrable
case (see text) following linear growth (dashed).
}
\label{fig:JCcorr}
\end{figure}

\begin{figure}
\centerline{\includegraphics[width=121mm]{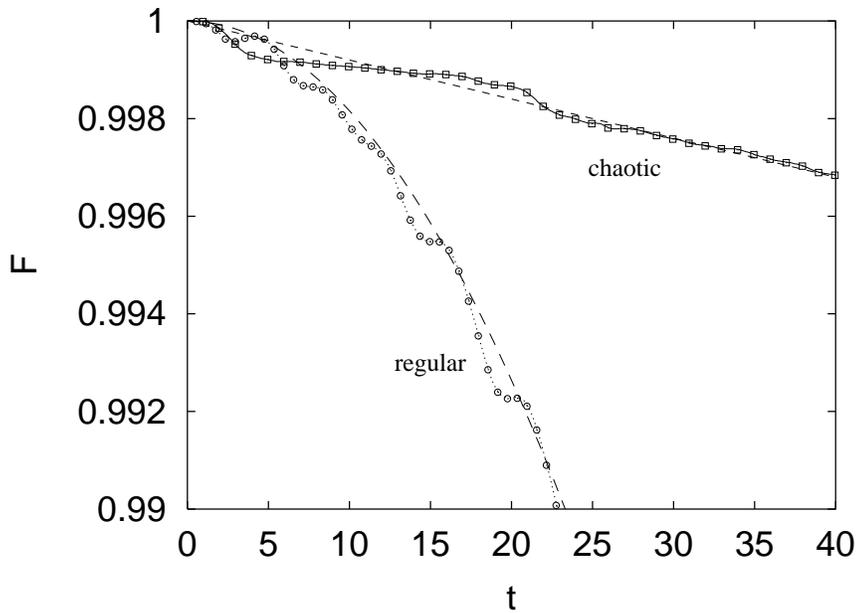}\hspace{8.5mm}}
\caption{
Fidelity decay of the coherent initial state in JC model for regular and
almost fully chaotic dynamics (see text).
Dotted and full curves correspond to numerically exact evaluations of fidelity,
whereas circles and squares reproduce a precise evaluation of the linear response formula (\ref{eq:F2nd})
in terms of quantum correlation functions (of fig.~\ref{fig:JCcorr}).
Dashed curves are best fitting linear and quadratic decays.
}
\label{fig:JCfidLR}
\end{figure}

As a second model we choose a Jaynes-Cummings (JC) Hamiltonian, a popular model in the realm
of quantum optics. JC model is an autonoumous (time-independent) system of 
an harmonic oscillator interacting with a rotor, {\it e.g.} one
mode of electromagnetic field, and a spin, say of an atom.
Suppose the oscillator is described by standard anihilation/creation operators
$a,a^\dagger$, and the spin of (half)integer value $J$ by SU(2) variables $J_{\pm},J_{\rm z}$.
Then the JC Hamiltonian reads
\begin{equation}
H = \hbar \omega a^\dagger a+ \hbar \epsilon J_{\rm z} + \frac{\hbar}{\sqrt{2J}} \left(G (a J_{+} + a^\dagger J_{-}) + G' (a J_{-} + a^\dagger J_{+})\right).
\end{equation}
Here the last (counter rotating) term has been included to allow for 
chaotic motion \cite{nemes}
Contrary to the kicked top model and the framework of previous theoretical discussion, 
time $t$ is here a continuous variable. There is, however, a natural limiting procedure in order
to obtain continuous time versions of results of sections 2 and 3 by replacing all the sums 
like $\sum_{t'=0}^{t-1}$ by integrals $\int_0^t dt'$. 
A convenient scaling towards the classical limit $J\to\infty$, $\hbar\to 0$, 
is obtained by fixing classical angular momentum $\hbar J = 1$, hence $\hbar=1/J$.
We perturb JC model by {\em detuning}, i.e. slightly changing the magnetic field parameter $\epsilon$,
so the generator of perturbation reads $V = \hbar J_{\rm z}$.
As an initial state we here choose only the coherent state, namely a direct product of oscillator coherent
state with complex parameter $\alpha$, and SU(2) coherent state with another complex parameter
$\tau=e^{i\varphi}\tan(\vartheta/2)$:  
\begin{equation}
\ket{\psi} = e^{\alpha a^\dagger - \alpha^* a}\ket{0}_2 \otimes
 (1+|\tau|^2)^{-J} e^{\tau J_{-}} \ket{J,J}_1.
\end{equation}
The classical JC Hamiltonian is integrable if either $G=0$ or $G'=0$,
whereas it can display all the variety of KAM regimes as both parameters $G$, 
$G'$ are increased and
reaching practical ergodicity for sufficiently large $G,G'$.
We study numerically two different sets of parameters:
In both we choose $\omega=\epsilon=0.3$, and
(i) $G=1,G'=0$ corresponding to classically {\em regular} motion,
and (ii) $G=G'=1$ corresponding to classically almost fully {\em chaotic} motion.
In both cases we choose the same initial coherent state with parameters 
$\alpha=1.15,\tau=0.29 + {\rm i}0.46$ and set angular momentum $J=4$.
We first choose a small perturbation parameter $\delta=0.005$.
Numerical results of time correlation functions and fidelity decay in such 
a linear
response regime are shown in figs.~\ref{fig:JCcorr},\ref{fig:JCfidLR}.

In order to reach beyond the linear response regime and go deeper into the semiclassical regime we use 
the kicked top model. There we may hope to exploit the behaviour of classical correlation functions in 
order to predict quantum fidelity decay. First, we turn to the regime of completely chaotic classical 
dynamics.

The classical correlation functions calculated by using the classical map (\ref{eq:KTclass}) are shown in 
fig.~\ref{fig:class30}. 
For $\alpha=30,\gamma=\pi/2$ the correlation function is oscillating with an 
exponential envelope hence the transport coefficient 
$\sigma_{\rm cl}=0.00385$ is quite small. 
For $\alpha=30,\gamma=\pi/6$ the correlation decay is 
monotonic and exponential with $\sigma_{\rm cl}=0.0515$. 
The decay of quantum fidelity (\ref{eq:Fmix}) can now be obtained by using the {\em classical} limit 
$\sigma\to\sigma_{\rm cl}$:
\begin{equation}
F_{\rm em}(t)=\exp{(-2\delta^2 J^2 \sigma_{\rm cl} t)}.
\label{eq:Fmclass}
\end{equation}
This formula has been compared with the exact numerical calculation of fidelity 
averaging over a set of random initial states.
We stress that we found {\em no observable difference} for sufficiently large $J$ when we have instead chosen 
a fixed coherent initial state, however this calculation is not shown in the figures.
As the finite size fidelity fluctuation level $F_{\rm ta}$ decreases with increasing Hilbert space dimension, 
we chose large $J=4000$ in order to be able to check exponential decay (\ref{eq:Fmclass}) 
over as many orders of magnitude as possible. 
\begin{figure}
\centerline{\includegraphics{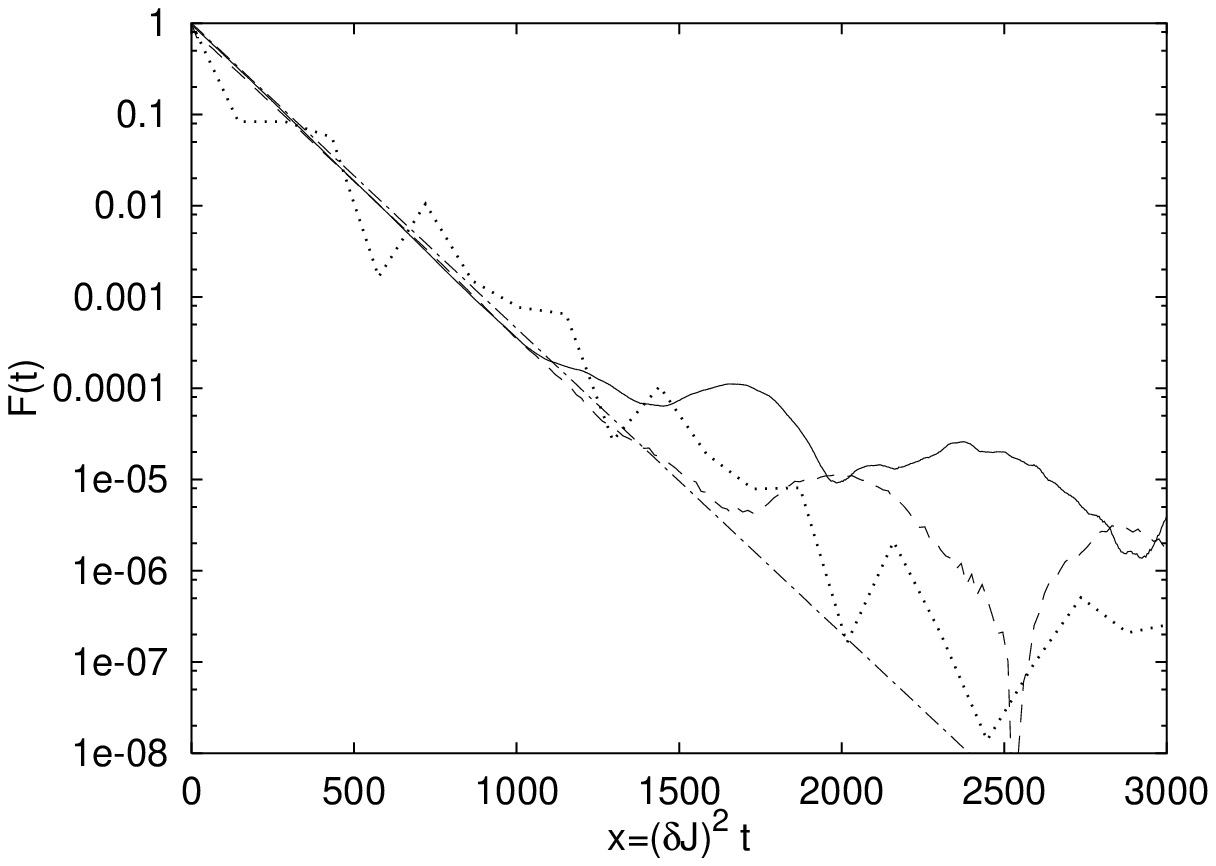}}
\centerline{\includegraphics{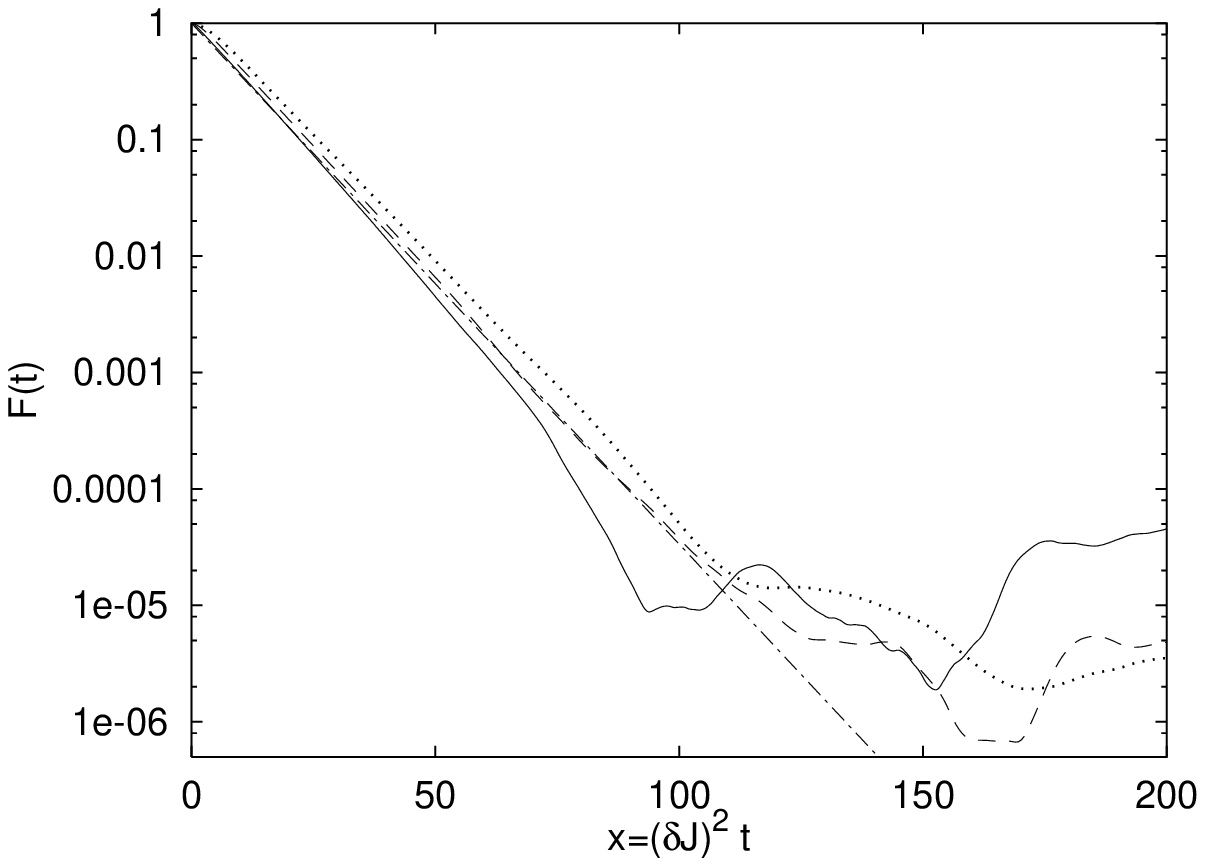}}
\caption{Quantum fidelity for a kicked top with parameters $\alpha=30$, $J=4000$ and full trace average. 
Top figure is for $\gamma=\pi/2$ and for $\delta=5\cdot 10^{-4},1\cdot 10^{-3},3\cdot 10^{-3}$ 
(solid, dashed, dotted curves, respectively)
Bottom figure is for $\gamma=\pi/6$ and $\delta=1\cdot 10^{-4},2\cdot 10^{-4},3 \cdot 10^{-4}$, 
(solid, dashed, dotted curves, respectively).
Chain line in both cases gives the theory (\ref{eq:Fmclass}) with classically computed $\sigma_{\rm cl}$.
Note that the largest $\delta=3\cdot 10^{-3}$ case in the top figure (dotted curve) corresponds to $\tau_{\rm}\approx 2$,
so it is already over the upper border of the regime ({\bf c}, subsect.~\ref{sec:timepert}) $\delta > \delta_{\rm mix}$ 
but the agreement with the theory (\ref{eq:Fmclass}) is still quite good, appart from oscillations. 
This is due to the oscillatory nature of time-correlations making the factorization assumption (\ref{eq:factor})
justified (on average) even for much smaller time $t$ as required.
}
\label{fig:j40k}
\end{figure}
The results are shown in fig.~\ref{fig:j40k}. The smallest and the largest $\delta$ shown, roughly correspond to borders 
$\delta_{\rm s}$ and $\delta_{\rm mix}$, respectively. As we can see, the agreement with an exponential decay is excellent, 
at least over four decades in the fidelity $F(t)$.
Note that for $\gamma=\pi/2$ and the largest $\delta=3\cdot 10^{-3}$ the time-scale of the decay of fidelity is comparable to the 
classical time-scale $t_{\rm mix}$ so the factorization assumption 
(\ref{eq:factor}) is strictly no longer applicable.
However, due to oscillatory correlation decay, overall agreement with the theory 
(\ref{eq:Fmclass}) is still rather good, but the
oscillations of the correlation decay are reflected by oscillations of the fidelity decay 
(around the theoretical exponential curve).
Of course one does not need such a large $J$ in order to have an exponential decay, but for smaller $J$ the fluctuation
level $F_{\rm ta}$ will be higher so the exponential decay (\ref{eq:Fmclass}) will persist for correspondingly smaller
time.
\par
\begin{figure}
\centerline{\includegraphics{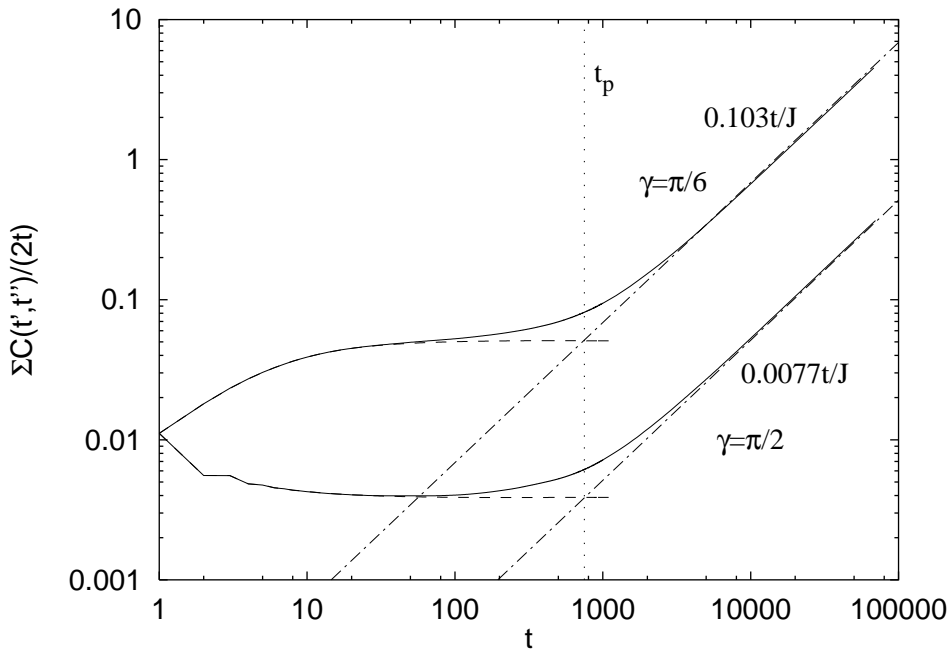}}
\caption{Finite time quantum correlation sum 
$\sigma(t)=\sum_{t',t''=0}^{t-1} {C(t',t'')/2t}$ (solid curves) 
and the corresponding classical sum 
$\sigma_{\rm cl}(t) = \sum_{t',t''=0}^{t-1}{C_{\rm cl}(t',t'')}/2t$ 
(dashed curves saturating at $\sigma_{\rm cl}$ and ending at $t \sim 1000$) for $\alpha=30,J=1500$. 
Upper curves are for $\gamma=\pi/6$ while lower curves are for $\gamma=\pi/2$. 
Chain lines are best fitting asymptotic linear functions corresponding to $\bar{C}t$, 
$0.0077t/J$ for $\gamma=\pi/2$ and $0.103 t/J$ for $\gamma=\pi/6$.}
\label{fig:cinf}
\end{figure}
Then we focus on the so-called perturbative regime $\delta < \delta_{\rm p}$ where the fidelity decay 
will be dictated by a 
finite size correlation average (\ref{eq:finitesize}), so according to eq. (\ref{eq:Fp})
\begin{equation}
F_{\rm pert}(t)=\exp{(-4 \delta^2 J \sigma_{\rm cl} t^2)}.
\label{eq:Fpert}
\end{equation}  
We numerically computed $\bar{C}$ (\ref{eq:Cinfty}) for $J=1500$, $\alpha=30$ in order to show that it is given by 
the theoretical value (\ref{eq:finitesize}). The quantum correlation function has been computed 
$C(t',t'') = \ave{\tilde{V}_{t'}\tilde{V}_{t''}}$ by means of a traceless perturbation
$\tilde{V} = \frac{1}{2}(J_z/J)^2 - \frac{1}{12}[(2J+1)(J+1)/J^2]\mathbbm{1}$.
In fig.~\ref{fig:cinf} we show a finite time correlation sum $\sigma(t)=\frac{1}{2t}\sum_{t',t''=0}^{t-1} C(t',t'')$ which
exhibits a crossover, at the Heisenberg time $t_{\rm p}=J/2$, from the plateau given by $\sigma_{\rm cl}$ to a 
linear increase $\bar{C}t$ due to finite size correlation average (\ref{eq:finitesize}) $\bar{C}=4 \sigma_{\rm cl}/J$.
\begin{figure}
\centerline{\includegraphics{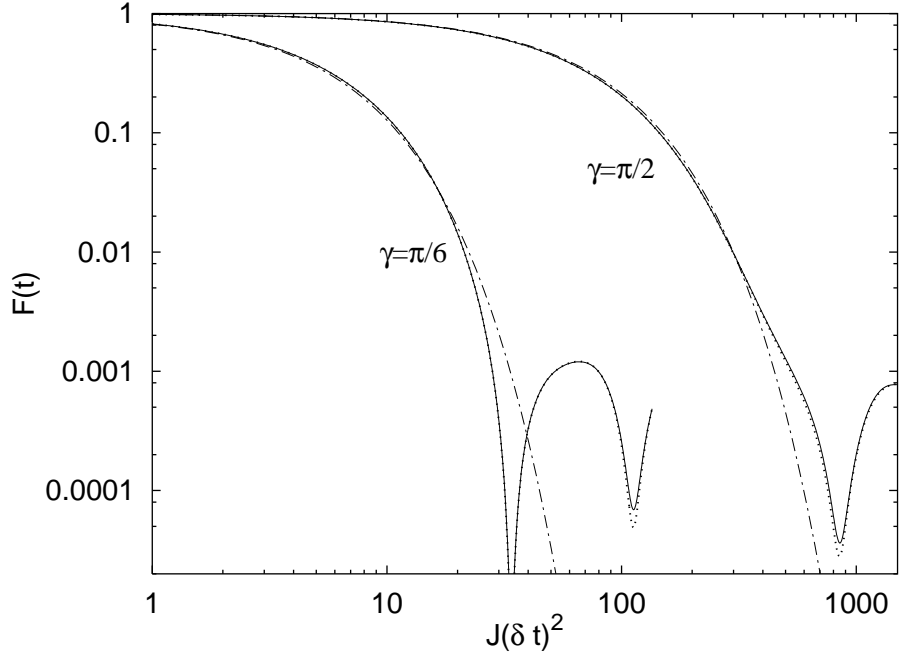}}
\caption{Quantum fidelity in the perturbative regime $\delta<\delta_{\rm p}$ for $\alpha=30$, $J=1500$, and 
$\gamma=\pi/2$ and $\pi/6$, calculated as a full trace Hilbert space average as a function of the scaled
variable $J(\delta t)^2$. For $\gamma=\pi/2$ data for $\delta=1\cdot 10^{-6}$ (solid curve) and $5\cdot 10^{-6}$ 
(dotted curve) are shown. For $\gamma=\pi/6$, $\delta=3\cdot 10^{-7}$ (solid) and $1\cdot 10^{-6}$ (dotted) are shown. 
Note that for both $\gamma$ the curves for both $\delta$ practically overlap. The chain curves are theoretical 
predictions (\ref{eq:Fpert}) with classically computed $\sigma_{\rm cl}$.}
\label{fig:j15kpert}
\end{figure}
The excellent agreement between prediction (\ref{eq:Fpert}) and full numerical calculation of fidelity is shown in 
fig.~\ref{fig:j15kpert}. 
In view of our findings this so-called \cite{Beenakker} perturbative regime can be understood as a simple consequence of a 
finite Hilbert space dimension. For times larger than the Heisenberg time 
$t_{\rm p}$ every quantum system behaves effectively as an integrable one, 
{\it i.e.} with a finite time average correlation plateau.

Next, we turn to the regime of nonergodic, say regular classical dynamics of 
the kicked top, which is
realized for  small value of $\alpha$. If the classical phase space has a mixed (KAM) 
structure, the non-mixing regime of fidelity decay may be obtained by choosing a localized initial state 
(e.g. coherent state) 
located in a regular part of the phase space. 
Such a situation may easily lead to the opposite conclusion (as compared to generic situation) 
for an insufficiently 
large dimension ${\cal N}$. As discussed in subsect.\ref{sec:timepert}, the fidelity fluctuation plateau 
is determined by 
the number of constituent propagator eigenstates $\ket{\phi_n}$ which are effectively needed to expand 
the initial state. 
For a coherent state sitting inside a (not too large) regular (KAM) island this number can be fairly small 
for numerically 
realizable Hilbert space dimensions, thus prohibiting any significant fidelity decay as observed in Ref.\cite{Peres2}
We would still see the initial quadratic decay in the linear response regime but we would not be able
to verify higher orders in the long-time expansion of fidelity.
In order to produce a situation numerically as clean as possible, we choose a small value of 
parameter $\alpha=0.1$, 
such that the classical dynamics is almost integrable and the majority of phase space corresponds 
to regular motion so that
the number of constituent eigenstates for coherent states is as large as possible (on average).
\par
Here we focus on the case $\gamma=\pi/2$. 
For small $\alpha$, the quantum and classical evolution is a (slightly perturbed) rotation around ${\rm y}$ 
axis and the time averaged perturbation can be computed analytically (to leading 
order in $\alpha$) as
\begin{equation}
\bar{V}=\frac{1}{4J^2}(J_{\rm z}^2+J_{\rm x}^2)=\frac{1}{4}\left(1-(J_{\rm y}/J)^2\right),\qquad
\bar{v}=\frac{1}{4}(1-y^2).
\label{eq:ave} 
\end{equation}
We will now use these approximate analytical results for $\alpha\to 0$ to compare with
numerics for $\alpha=0.1$. Note that our leading order analytical approximations could easily be systematically
improved using a classical perturbation theory (treating $\alpha$ as a perturbing parameter).
Since the agreement, as shown below, is almost perfect in all cases, we see no need for refinement at this 
level.
\par
First consider a random initial state.
Starting from expression (\ref{eq:Favg}), $F(t)=|\ave{\exp{(i t \bar{V}\delta/\hbar)}}|^2$,
write the fidelity as a sum over all 
eigenvalues of $J_{\rm y}^2$, namely $(2m-1)^2$, for $m=1,\ldots,J/2$ (in OE subspace),
\begin{equation}
F(t)=\left|\frac{2}{J} \sum_{m=1}^{J/2}{\exp{(i \delta t (2m-1)^2/4J)}}\right|^2.
\label{eq:Fexact}
\end{equation}
For large $J$ we can replace the sum with an integral and get
\begin{equation}
F(t)= \frac{\pi}{\delta J t} \left|{\rm erfi}(\frac{1}{2} e^{i \pi/4} \sqrt{\delta J t})\right|^2,
\label{eq:Ferfi}
\end{equation}
\begin{figure}
\centerline{\includegraphics{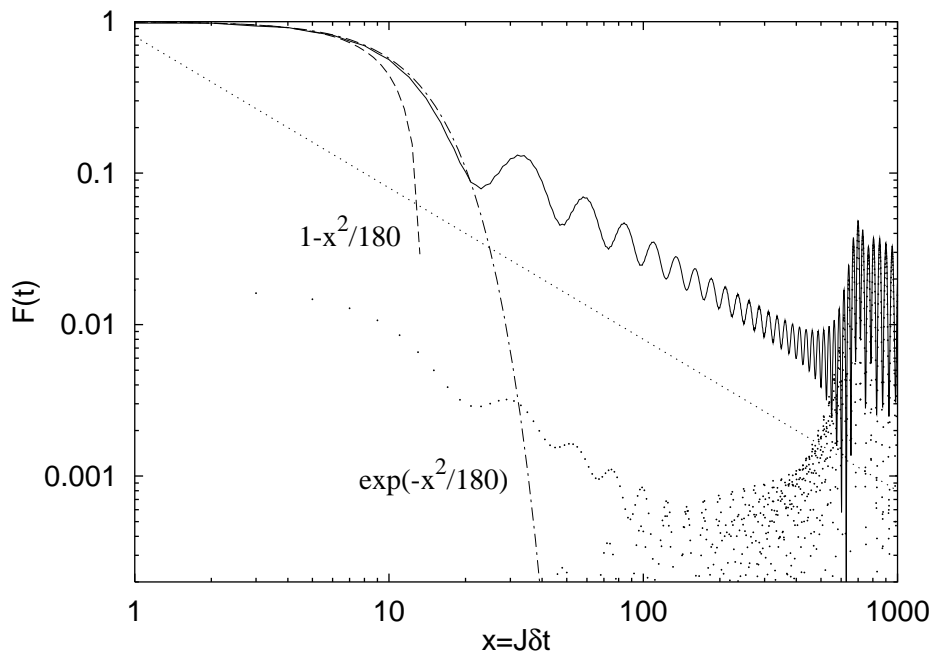}}
\caption{Fidelity in the near-integrable regime for $\alpha=0.1$, $\gamma=\pi/2$, $\delta=0.01$, $J=100$, and 
$\rho=\mathbbm{1}/J$, in the OE subspace. 
Solid curve gives the result of a numerical simulation. Isolated dots denote differences between numerical 
calculation and the 
analytic formula (\ref{eq:Ferfi}) for $\alpha\to 0$ 
$|F_{\rm num.}(t)-F_{\rm anali.}(t)|$. 
The dotted line gives the predicted asymptotic decay $\propto t^{-1}$, and the dashed/chain curves are the 
predicted fidelity decays at small times, namely the second order expansion $F(t)=1-(Jt\delta)^2/180$, and 
the 'improved' Gaussian result $F(t)=\exp(-(Jt\delta)^2/180)$.}
\label{fig:analit}
\end{figure}
where ${\rm erfi}(z)=\frac{2}{i\sqrt{\pi}}\int_0^{iz}{e^{-t^2} dt}$ is a complex error function with an 
asymptotic limit 
$\lim_{x \to \infty}{|{\rm erfi}(\frac{1}{2} e^{i \pi/4} \sqrt{x})|}=1$ to which it approaches by oscillating 
around $1$. We thus have an analytic expression for the fidelity (\ref{eq:Ferfi}) in the case of an uniform 
average over the Hilbert space or, equivalently, for a random initial state. 
Its asymptotic decay is $t^{-1}$ which 
agrees with the general semiclassical asymptotics (\ref{eq:Fsqrt}). We expect initial quadratic decay 
(\ref{eq:Fr2}) for small times $t < \tau_{\rm ne}$. The
decay rate $\tau_{\rm ne}$ is determined by the time averaged correlation $\bar{C}$ (\ref{eq:Cinfty}) which can be
calculated explicitly in the limit $\alpha\to 0$ where the classical correlation function $C_{\rm cl}(t)$ alternates 
for even/odd times as
$C_{\rm cl}(2t)=\ave{\tilde{z}^2(0) \tilde{z}^2(2t)}/4=-1/90$, 
$C_{\rm cl}(2t+1)=\ave{\tilde{z}^2(0) \tilde{z}^2(2t+1)}/4=1/45$,
giving
\begin{equation}
\bar{C}_{\rm cl}\vert_{\alpha=0} = 
\frac{1}{2}\left(-\frac{1}{90}+\frac{1}{45}\right)=\frac{1}{180}.
\label{eq:Cinftykt}
\end{equation}
Fidelity is expected to decay as (\ref{eq:Fr2}) with 
$\tau_{\rm ne} = \sqrt{180}/(J\delta)$,
for short times, $t < \tau_{\rm ne}$. The short-time formula (\ref{eq:Fr2}) and the full analytic 
expression (\ref{eq:Ferfi}) 
are compared with the numerical simulation in fig.~\ref{fig:analit}. The agreement is very good and, 
surprisingly enough, 
the Gaussian approximation $F(t) = \exp(-(t/\tau_{\rm ne})^2)$ for small times 
is observed to be valid considerably beyond the second order expansion (\ref{eq:Fr2}). 
Quite interesting is the regime where the decay time $\tau_{\rm ne}=\sqrt{180}/(J\delta)$ for a ``regular'' 
dynamics with a random initial state will be smaller than the decay time 
$\tau_{\rm em}=1/(\delta^2 J^2 \sigma_{\rm cl})$ (\ref{eq:Fmclass}) for a ``chaotic'' dynamics. 
This will happen for $\delta <  1/(J \sigma_{\rm cl} \sqrt{180})$.
This border has the same scaling with $J$ as $\delta_{\rm mix}$ (\ref{eq:deltam}).    
\par
\begin{figure}
\centerline{\includegraphics{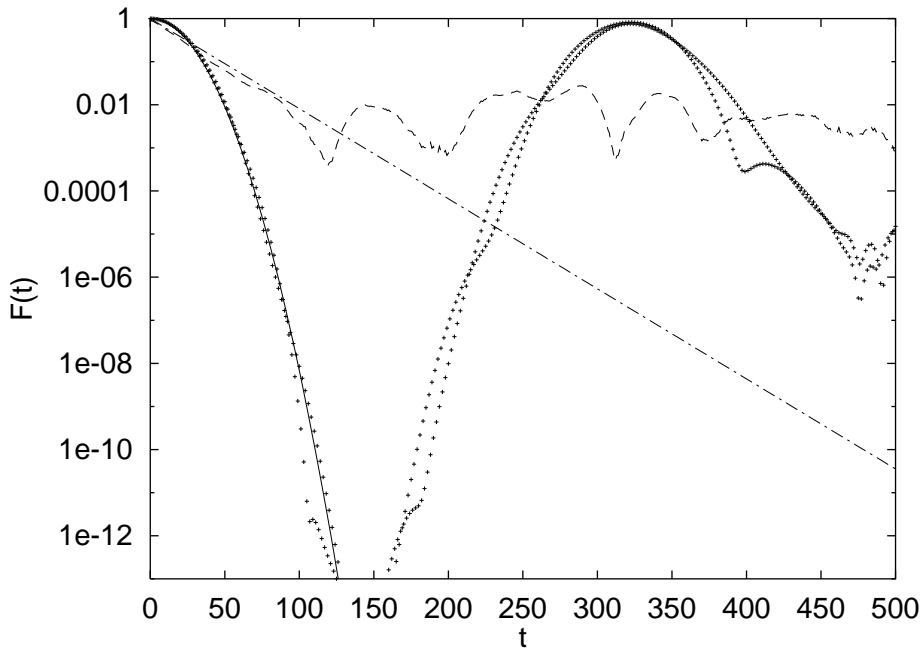}}
\caption{Fidelity for $\gamma=\pi/2$, $\delta=0.025$ and $J=100$ on the OE subspace.
The dashed curve is a simulation for $\alpha=30$ (mixing regime, full trace average). 
The pluses are for a pure coherent initial state (see text for details) at $\alpha=0.1$ (non-mixing regime). 
The chain and solid curves are, respectively, the theoretical exponential (\ref{eq:Fmclass}) 
and Gaussian (\ref{eq:Fktcoh}) decays.}
\label{fig:parad1}
\end{figure}

Second we consider  an SU(2) coherent initial state (\ref{eq:SU2coh}).
We could perform an exact analytical calculation for the fidelity decay in this particular case.
Rather than performing this calculation, we will illustrate the usefulness 
of a semiclassical formula for $F_{\rm ne}^{\rm coh}(t)$ (\ref{eq:Fcoh}). This is a more general approach, 
as an explicit 
analytical calculation is usually not possible. Let us denote by $\tilde{\vartheta},\tilde{\varphi}$ 
the spherical angular
coordinates measured with respect to the y-axis. Then $(I=\cos\tilde{\vartheta}=y,\tilde{\varphi})$ 
represent canonical 
action-angle coordinates for the integrable case $\alpha\to 0$. Furthermore, the coherent 
state (\ref{eq:SU2coh}) acquires 
a semiclassical Gaussian form (\ref{eq:CS}) in the EBK basis $\ket{n}$,
$J_{\rm y}\ket{n} = n\ket{n}$, namely
\begin{equation}
|\braket{n}{\tilde{\vartheta},\tilde{\varphi}}|^2 
\propto \exp{\left(-\frac{(n\hbar-\cos{\tilde{\vartheta}})^2}{\hbar\sin^2{\tilde{\vartheta}}} \right)},\quad \hbar=\frac{1}{J}.
\end{equation}
The squeezing parameter $\Lambda$ reads 
\begin{equation}
\Lambda=1/\sin^2{\tilde{\vartheta}}=1/(1-y^2).
\end{equation}
In order to apply the general formula (\ref{eq:Fcoh}) we need to express the classical time average (\ref{eq:ave}) 
in terms of a canonical action, $\bar{v}(I) = (1-I^2)/4$, and evaluate the derivative,
$|\partial \bar{v}(I)/\partial I|^2=\frac{1}{4} I^2$, giving
$F^{\rm coh}_{\rm ne}(t)=\exp{(-\delta^2 J t^2 I^2 (1-I^2)/8)}$. Rewriting this expression in 
terms of original spherical
angles $\vartheta$ and $\varphi$ measured with respect to ${\rm z}$-axis, we obtain
\begin{equation}
F^{\rm coh}_{\rm ne}(t)=
\exp{\left(-\frac{\delta^2 J t^2}{32} 
\{\sin^2{2\vartheta}\sin^2{\varphi}+\sin^4{\vartheta} \sin^2{2\varphi} \}
\right)}.
\label{eq:Fktcoh}
\end{equation}
We checked this formula by means of a numerical simulation and the result for a coherent state centered at 
$(\vartheta,\varphi)=\pi(1/\sqrt{3},1/\sqrt{2})$, 
for which $\sin^2{2\vartheta}\sin^2{\varphi}+\sin^4{\vartheta} \sin^2{2\varphi}=0.96$, is shown 
in fig.~\ref{fig:parad1}.
We can see that for $t > 50$ the fidelity in a non-mixing regime ($\tau_{\rm ne-coh}=23$) is lower 
than the fidelity in a mixing regime ($\tau_{\rm em}=42$). For larger times, $t > t^*$, 
non-mixing decay $F^{\rm coh}_{\rm ne}$ 
displays revivals of fidelity which are a simple manifestation of a {\em beating} phenomenon
due to sligtly different frequencies of revolution of wave-packets along unperturbed and perturbed KAM
tori. Note that this effect of revivals is drastically reduced in more than one dimension, $d\ge 2$,
where there are typically several different incomensurate frequencies \cite{ktop}.

\begin{figure}
\centerline{\includegraphics[width=4.5in]{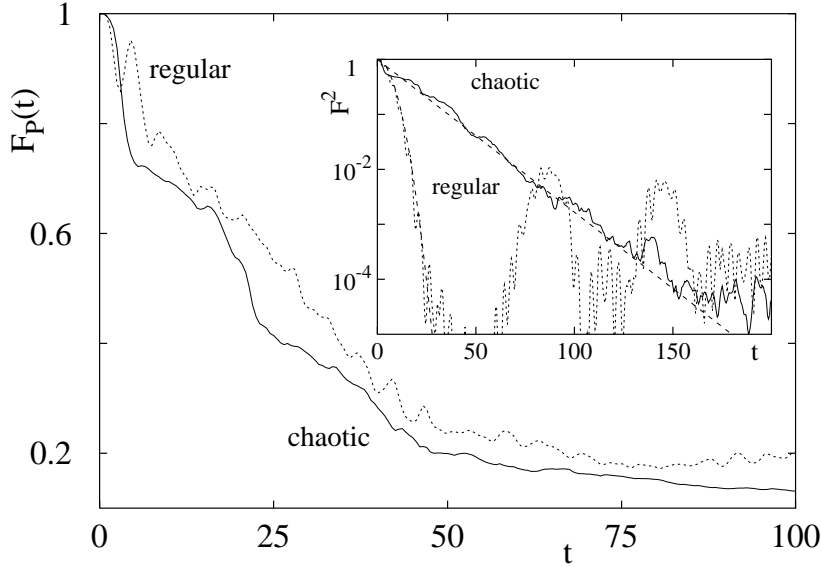}}
\caption{Purity fidelity $F_{\rm P}(t)$ (main figure) and squared fidelity $[F(t)]^2$ (inset) 
in chaotic regime (solid curves) and in integrable regime (dotted curves), for $\delta=0.1$.
The dashed lines indicate the linear and quadratic approximation respectively.
Note the differences in vertical scales.
}
\label{fig:purf}
\end{figure}

Next we make numerical experiments on the decay of purity fidelity. Again we choose the JC 
Hamiltonian and treat the oscillator and the spin as two parts of a composite system.
We first report a calculation with a strong perturbation  $\delta=0.1$, that rapidly exceeds  the 
realm of validity of linear response, in Fig.~\ref{fig:purf} where the main figure gives the 
purity fidelity and the inset the fidelity for comparison. For the fidelity decay (inset) we find excellent 
agreement with exponential decay (\ref{eq:Fmix}) in a chaotic 
regime and with a faster Gaussian decay in a regular regime (\ref{eq:Fcoh}), where the 
decay rates are fixed as above.
For purity fidelity, however, we find 
already at $t \approx 20$, that the decay starts to be influenced by the saturation value of 
$F_{\rm P}(t \to \infty) \approx 1/(2J+1)$. Therefore purity fidelity is
higher for the integrable case than for the chaotic one  not only at short times, as expected, but
even at large times. This is relevant, because we shall next choose a weak perturbation $\delta=0.005$
to avoid this problem. 
\begin{figure}
\centerline{\includegraphics[width=4in]{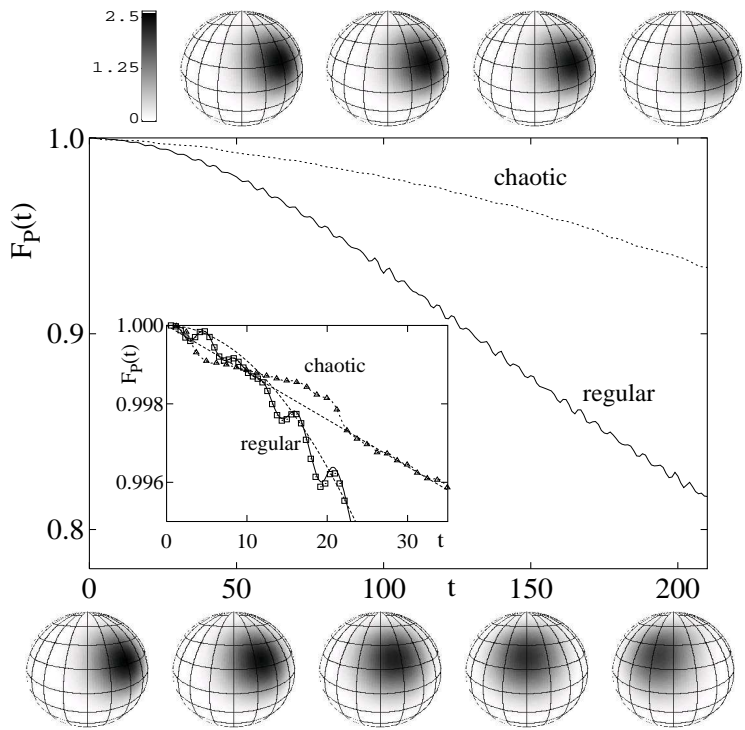}}
\caption{Echo dynamics for weak coupling: $\delta=0.005$.
 Square of the Wigner function for chaotic dynamics (top diagrams) and integrable dynamics as a function of 
time (bottom diagrams) at times corresponding to the axis.
Grey code: top left.
Purity fidelity is shown in the frame on the same time scale and for short times in the inset. 
Full curves show the complete numerics, 
symbols the evaluation starting from the numerical correlation functions of Fig.~\ref{fig:JCcorr} and dashed curves 
the linear or quadratic approximation.}
\label{fig:wig}
\end{figure}
We expect and find the crossover after a fairly short time.
This calculation allows comparison with theory as well as an illustration 
of the evolution of the square of the Wigner function, corresponding to the reduced density matrix
$\rho_1(t)$ for the angular momentum  states on the sphere 
using the definition of Ref.\cite{Agarwal81}. 
Near the top and bottom of Fig.~\ref{fig:wig} 
we see this evolution for the chaotic and the integrable Hamiltonian 
respectively. 
In the center of the figure we plot the purity fidelity on the same time scale as the Wigner 
functions in the main frame and an amplification of short times in the inset.
Again, we note faster decay of purity fidelity in the case of regular dynamics as compared to 
chaotic dynamics.
We observe detailed agreement of numerics with results obtained from the numerical values of
the correlation integrals (\ref{eq:F2nd},\ref{eq:Fp2nd}) reproducing the 
oscillatory structure. From the same correlation integrals we obtained the coefficients 
for the linear and quadratic decay, which  agree well if we discard the oscillations. 
We see a crossing of the two curves at $t=t^*_{\rm P}\approx 12$ 
for $F_{\rm P}$.
These times are larger than the Zeno time ($\approx 1$) and indicate 
the competition of the decay rate and the decay shape as expected for a non-small value of $\hbar=1/4$.
It is important to remember that the integral over the square of the Wigner function gives the purity
and therefore the fading of the picture will be indicative of the purity decay. On the other hand the 
movement of the center is an indication of the rapid decay of fidelity 
(not shown in the figure).

\section{Summary and discussion}

In this paper we have outlined a simple theory of stability of quantum time evolution with respect to small
variation of the Hamiltonian or, more generally, of the time-evolution propagator.
The central object of study is the fidelity, a cross correllation between two  states originating
from the same initial state but subject to two slightly different time evolutions.
An alternative interpretation of the whole theoretical setup can be given in terms of quantum dynamical
irreversibility in the Loschmidt echo gedanken experiment.
This intepretation allows even for more general characterization of the quality of quantum echo,
as given for example by purity fidelity which measures the quality with which the final state,
after an echo experiment, factorizes in the direct product Hilbert space of a composite system.
When one part of a composite system is interpreted as environment, then the decay of purity fidelity can
be directly linked to the growth of decoherence.

The main result reported here is a simple linear response (or Kubo-like) formula 
(\ref{eq:F2nd}) which relates the fidelity decay, and quite similarly also the purity fidelity decay, 
to the total sum (or integral) of two point time
autocorrelation function of the perturbation.
In the limit of infinite Hilbert space dimension we have found 
{\em exponential} fidelity decay on a time-scale $\tau_{\rm em} \propto (\hbar/\delta)^2$, for 
{\em quantum ergodic and mixing} systems, whereas for {\em non-ergodic} systems we have found much faster 
decay (in the sufficiently `quantum' regime where $\delta \ll \hbar$, but note that both, $\delta$ and $\hbar$ are dimensionless parameters) on a time-scale 
$\tau_{\rm ne} \propto \hbar/\delta$ for random initial states or $\tau_{\rm ne-coh} \propto 
\hbar^{1/2}/\delta$ for coherent initial states (minimal uncertainty wavepackets)
where the fidelity $F(t)$ is given by a {\em Fourier transformation of the 
local density of states of the time averaged perturbation operator}.
A special emphasis has been given to the semiclassical theory of fidelity of small but 
finite values of $\hbar$, 
where different regimes and the corresponding time and perturbation scales are carefully discussed, 
and where fidelity decay may asymptotically (as $\hbar\to 0$) be evaluated in terms of classical quantities only. 
Interestingly, finite size fluctuations of fidelity (for very long times at a finite Hilbert space dimension) 
have been shown to be given by the {\em inverse participation ratio} of the eigenstates of the perturbed 
evolution operator in the eigenbasis of the unperturbed propagator. 
The surprising aspects of our relations are basically due to non-interchangability of the limits $\delta\to 0$ and 
$\hbar\to 0$ as the relevant decay time-scales are only functions of the ratio $\delta/\hbar$.
Therefore a different and intuitively expected behaviour, namely faster fidelity decay for mixing than regular 
dynamics, is obtained in the `classical' regime where $\hbar \ll \delta$ 
(or making the limit $\hbar\to 0$ prior to $\delta\to 0$).
We should note that a similar, reassuring result has been found by applying our formalism to inspect the analogous classical fidelity\cite{ktop,benenti,veble} for the unitary 
Perron-Frobenius evolution of volume (area) preserving maps \cite{ktop}: 
there the classical fidelity for regular dynamics has been found to decay on a time-scale $\propto \delta^{-1}$,
{\em i.e.} the same as for quantum fidelity decay of regular dynamics, 
whereas for a chaotic dynamics, the fidelity decay is governed by
a maximal Lyapunov exponent $\lambda$ on a short time-scale 
$\propto \ln(1/\delta)/\lambda$.

Our findings may have some important implication in the understanding of the
stability of quantum computation \cite{QC} where different regimes and time-scales \cite{DimaKR} 
should perhaps be explained in terms of intrinsic dynamics of a particular quantum algorithm. 
On the other hand, our results may also shed new light on the relation
between decoherence and dynamics \cite{Zurek}. In particular, due to the quite special role of 
coherent initial states on the short Ehrenfest time-scale one may expect the results to be quite 
different for a random initial state and/or for longer time-scales (see \cite{Gorin}).

\section*{Acknowledgements}

The work has been supported by the Ministry of Education, Science and Sport of Slovenia,
and in part by the grants: IN-112200, DGAPA UNAM, Mexico,  25192-E CONACYT Mexico, and
DAAD19-02-1-0086, ARO United States.

\end{document}